# How pulse polarity and photoionization control streamer discharge development in long air gaps

A Yu Starikovskiy[1,*] and N L Aleksandrov[2]

[1]Princeton University, Princeton, NJ08544, USA
[2]Moscow Institute of Physics and Technology, Dolgoprudny, 141700, Russia
[*]Author to whom any correspondence should be addressed
E-mail: astariko@princeton.edu

**Abstract**

A 2D approach is used to simulate the properties of positive and negative streamers emerging from a high-voltage electrode in a long (14 cm) air gap for standard pressure and temperature. The applied voltage varies from 100 to 500 kV. To reveal the influence of photoionization, the calculations are made for various rates of seed electron generation in front of the streamer head. The difference between the properties of positive and negative streamers is associated with the different directions of the electron drift ahead of the streamer head. As a result, the peak electric field at the streamer head and the streamer velocity are higher for positive voltage polarity. The average electric field in the negative streamer channel is approximately twice that in the positive streamer channel, in agreement with available measurements in long air gaps. It is shown that photoionization in front of the streamer head is important not only for the development of strong positive discharges, but for the development of strong negative discharges as well. An increase in the photoionization rate increases the propagation velocity of the positive streamer and retards the propagation of the negative streamer.

# Introduction

Streamers are experimentally studied almost over a century because streamer discharges are an important stage of electrical breakdown in gases, including discharge phenomena in the atmosphere (lightning, sprites and blue jets) [1-4]. Streamer coronae near the wires of transmission lines lead to considerable energy losses. On the other hand, using streamer discharges, it is easy to generate non-equilibrium atmospheric pressure plasmas. Therefore, streamers are often met in plasmachemical systems, among them ozone generators and plasma generators for air purification and plasma medicine [5-7].

Experiments have provided a large amount of information on the development of a solitary streamer discharge and streamer flashes in which tens and hundreds of streamer channels propagate simultaneously. The properties of streamer discharges depend on the polarity of applied voltage. The threshold voltage for initiation of positive streamers is much lower than that for initiation of negative streamers [1, 8]. Positive streamers are faster and their channels are thicker in comparison with negative streamers, at least in short (4 cm) air gaps [8]. The average electric field in the negative streamer channel required for bridging long air gaps under standard atmospheric conditions is approximately twice that in the positive streamer channel. This field is in the range 10-16 kV/cm for negative polarity and in the range 4.5-5 kV/cm for positive polarity [1]. This difference leads to a practically important conclusion: for the same value of applied voltage, positive streamers can bridge air gaps which are twice longer than the gaps covered by negative streamers. In addition, because of the difference in the average electric field in streamer channels, the streamer zone of a long negative leader is much more complicated in its structure in comparison with the streamer zone of a positive leader [1, 2]. As a result, negative leaders develop in a stepped way. Negative stepped leaders have been repeatedly observed in laboratories and during the development of a downward lightning discharge under thunderstorm conditions [1-4].

Analytical and numerical studies of streamer discharges allowed an understanding of the dominant mechanisms of discharge development and gave an insight into the relationships between the fundamental characteristics of electron-molecule interactions and the properties of streamer plasmas. Most of analytical studies did not address the effect of polarity on streamer properties or considered the development of positive streamers only [1, 2, 9]. A simultaneous analysis of positive and negative streamers was analytically made only for short (≈ 1 cm) air gaps [10, 11]. The focus of numerous numerical studies was on the properties of practically important positive streamers. Comparative analysis of positive and negative streamers has been numerically made only for the discharges developing in short (1 cm) atmospheric pressure air gaps [10, 12]. In [10], streamer propagation in weak uniform electric fields was simulated, whereas Luque et al. [12] addressed both streamer discharges in uniform electric fields and streamer development from a high-voltage electrode in a non-uniform electric field produced by a moderate (20 kV) pulsed voltage. It was shown that the peak electric field at the streamer head is higher for positive voltage polarity [10]. As a result, the plasma density is higher in the channels of positive streamers. The electric field in the streamer channel was lower for positive polarity;

however the effect of polarity on the electric field in the channel was small. Positive streamers could be faster or slower than negative streamers. It is important that the difference between the properties of positive and negative streamers decreased with the external electric field increase. Numerical simulations [12] showed that the outward drift of seed electrons in front of the negative streamer head leads to a smaller field enhancement at the head and ultimately to a slower propagation of negative streamers, in agreement with observations [8].

The difference in the behavior of positive and negative streamers also follows from the simulations of double-headed streamers in weak homogeneous electric fields under low-pressure conditions of sprite discharges [13, 14]. As for the atmospheric pressure simulations [12], positive streamers under sprite conditions were characterized by higher peak electric fields in the streamer head, higher plasma densities in the streamer channels and higher velocities of propagation. In addition, the channels of positive streamers were thicker than those of negative streamers.

Particular attention has been given to the role of seed electrons in the development of streamer discharges. These electrons are usually generated in front of the streamer head due to photoionization by UV photons originating from a region of high electric field in the streamer head. A key role of this process in the propagation of positive streamers is beyond question because they develop opposite to the electron drift. The importance of the photoionization process on the seed electron production for negative streamers is under discussion. It is assumed that photoionization is less important for negative streamer development in comparison with positive streamers [1]. According to simulations [12, 15], photoionization slightly affects the properties of negative streamers developing in weak electric fields, whereas the influence of photoionization becomes much more profound in strong electric fields. It follows from the simulation [16] that negative streamers fail to develop in the absence of photoionization and other volumetric mechanisms of seed electron generation. All these simulations were made for short (1 cm) air gaps. Besides, great differences between the streamer properties under sprite conditions and in atmospheric pressure air are largely explained by a more efficient photoionization in low-pressure air due to a reduced collisional quenching of electronically excited singlet states of $N_2$ molecules, which emit UV photons producing seed electrons ahead of streamer heads [13, 14].

Previous studies of short streamers showed that the effect of polarity on streamer discharges becomes less profound with increasing external electric field and applied voltage. Therefore, it is important to comparatively study the properties of positive and negative streamers developing in long air gaps when high voltages are applied. Such situations happen during the initiation and development of a leader discharge on a laboratory scale and during the progression of lightning leaders in the cloud-to-ground gaps [1, 2]. Many characteristics of a leader discharge are controlled by the processes in the streamer zone that always exists in front of the leader channel [1]. This paper addresses the effect of polarity on the properties of a streamer discharge emerging from a high-voltage electrode in a long (14 cm) air gap. For this purpose, we made a 2D numerical simulation of positive and negative streamers at voltages in the range 100-500 kV under standard atmospheric conditions. To show the role of photoionization in the discharge development, calculations were repeated for various efficiencies of photoionization in front of the streamer head. An analytical approach was also used to qualitatively analyze streamer properties for different voltage polarities and photoionization rates.

## Simulation model

The initiation and propagation of streamers were investigated using an axisymmetric two-dimensional hydrodynamic model. The model and computational method were presented in detail in previous papers [9, 17-20]. We considered a single streamer developing in a 14 cm plane-to-plane gap. The streamer was initiated near the high-voltage electrode and propagated along the axis of the gap to the grounded electrode. All calculations were made for air at a pressure of 760 Torr. The initial gas temperature was 300 K. Figure 1 demonstrates typical results of calculations of the electron density and electric field for positive and negative streamers. The contours show the distributions for $t = 3.5$ ns at $U = +400$ kV and $t = 4.1$ ns at $U = -400$ kV.

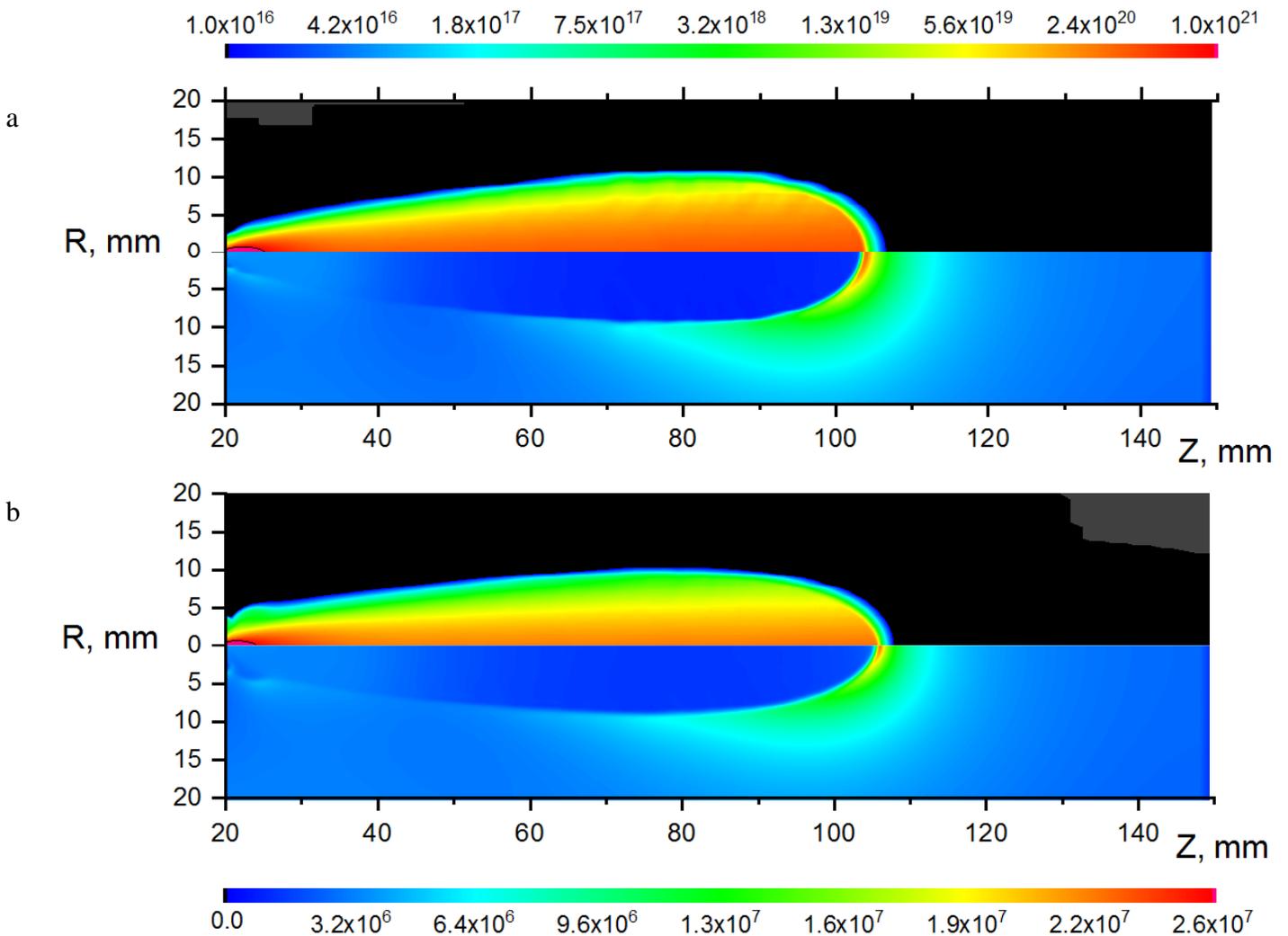

Figure 1. Contours of (top) electron density [1/m$^3$] and (bottom) electric field [V/m] for (a) positive and (b) negative streamers. Calculations are made for $t = 3.5$ ns at $U = +400$ kV and $t = 4.1$ ns at $U = -400$ kV. The top and bottom color scales correspond to electron density and electric field, respectively.

In both cases, the streamers reached approximately the same length, which indicates that the positive streamer propagates faster than the negative one. The electric field in front of the streamer head and the electron density in the channel are also higher for the positive streamer (figure 1).

The computational domain was 150×150 mm$^2$ (figure 2). The high-voltage electrode was a combination of a plate at Z = 10 mm with a semi-ellipsoidal needle (large semi-axis 8 mm and small semi-axis 0.8 mm) protruding from the center of the plate. Thus, the discharge gap had an asymmetric geometry, which guaranteed that the streamer emerged from the high-voltage electrode. The computational mesh was adapted as the peak of the electric field moved along the discharge gap. The typical number of cells was $N_z \times N_r = 1024 \times 256$. The minimum spatial step in the radial direction was $\delta r_{min} = 5$ μm, and further increased exponentially to $\delta r_{max} \cong 1$ cm at $R_{max} = 150$ mm. In the axial direction, the uniform mesh with a step of $\delta z_{min} = 5$ μm occupied a region of $\Delta z = 2$ mm after which the space step increased exponentially as it approached the boundary of the computational domain. The change in the spatial step size between adjacent cells was limited to 10% in order to avoid the increase of the computational errors on an inhomogeneous grid. Such a mesh was shifted each time the peak of the electric field moved to a distance of 0.5 mm. As a result, the regions of intense ionization and photoionization in front of the streamer head were always calculated on a homogeneous grid of 400×200 cells with the space step $\delta r_{min} = \delta z_{min} = 5$ μm.

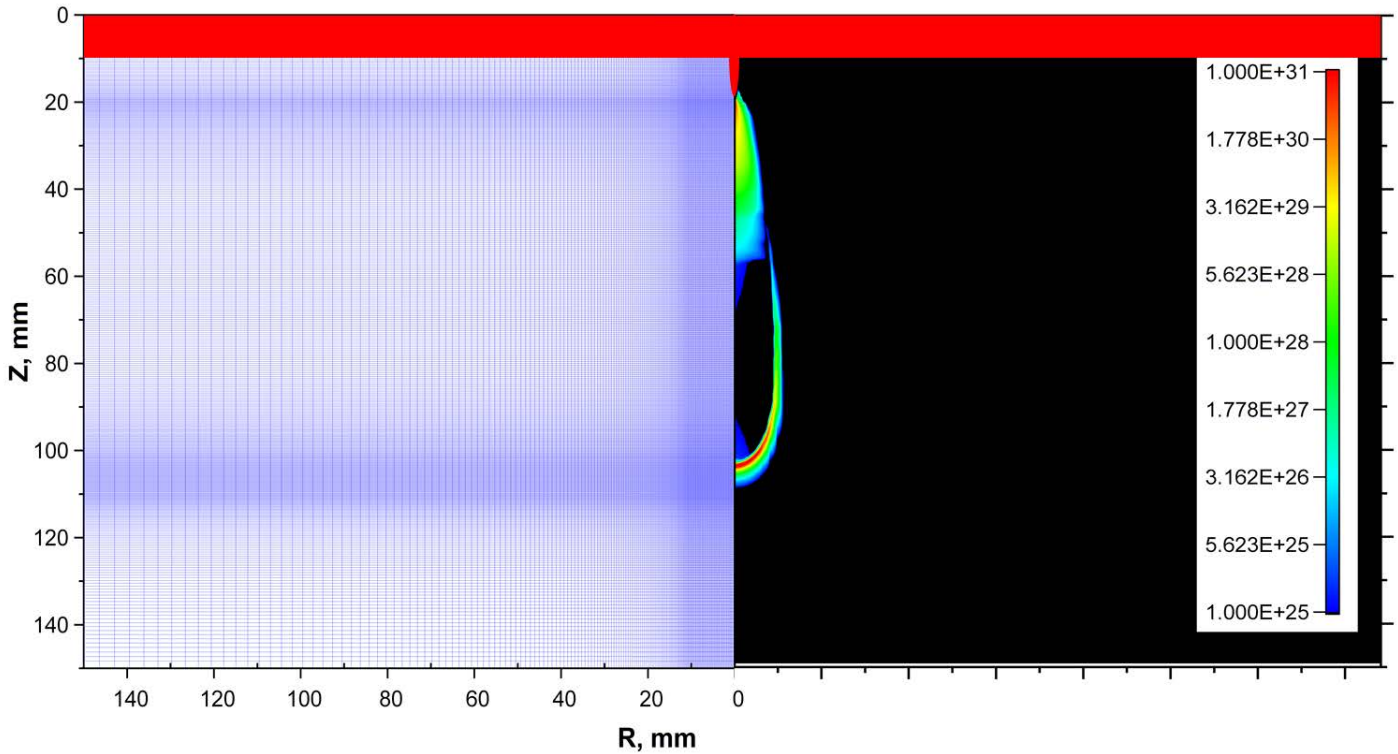

Figure 2. Left: the geometry of the discharge gap and the adaptive computational grid (each 4$^{th}$ cell is shown). High-voltage electrode is located on the top and marked by red. Right: contours of the ionization rate distribution $dn_e/dt$ in [m$^{-3}$/s] 3.5 ns after the onset of the +400 kV high voltage pulse. The grounded electrode is located at Z = 150 mm.

The time step was, as a rule, around $\delta t \cong 5\times10^{-14}$ s, which was required for the stability and accuracy of the numerical scheme. We assumed that the applied voltage increased linearly to the maximum value for 1 ns and remained constant for $t > 1$ ns.

The transport and reactions of charged particles were considered in the local approximation, while the non-local approach was used to describe photoionization generating seed electrons in front of the streamer head by means of ionizing radiation emitted from the head and streamer channel:

$$\frac{\partial n_e}{\partial t} + \text{div}(\vec{v_e} \cdot n_e) = S_{ion} + S_{photo} - S_{att} - S_{rec}^{ei} \quad (1)$$

$$\frac{\partial n_p}{\partial t} = S_{ion} + S_{photo} - S_{rec}^{ei} - S_{rec}^{ii} \quad (2)$$

$$\frac{\partial n_n}{\partial t} = S_{att} - S_{rec}^{ei} \quad (3)$$

$$\Delta\varphi = -\frac{e}{\varepsilon_0}(n_p - n_e - n_n) \quad (4)$$

where $S_{ion}$ is the ionization rate, $S_{photo}$ is the rate of photoionization, $S_{att}$ is the rate of electron attachment, and $S_{rec}^{ei}$ and $S_{rec}^{ii}$ are rates of electron-ion and ion-ion recombination. The system of equations (1) - (3) was converted into a finite-difference form by the control volume method. Equations (1) and (2) were solved by splitting between physical processes. The transport of charged particles was calculated using an explicit finite-difference backward approximation of the first order of accuracy in space and time, and the right sides of the equations were calculated using the Euler technique of the first order in time [9]. To solve the Poisson equation (4), the Gauss–Seidel overrelaxation technique was used.

The photoionization source term in equations (1)-(2) was calculated using the approach [21]:

$$S_{photo} = \frac{1}{4\pi} \frac{p_q}{p+p_q} \int_V d^3 r_1 \frac{S_{ion}(r_1)}{|r-r_1|^2} \Psi(|r-r_1|p), \quad (5)$$

where $p$ is the gas pressure, $p_q$ is the quenching pressure ($p_q = 30$ Torr for air [21]), $\Psi(|r-r_1|p)$ is the coefficient of absorption for the ionizing radiation in the gas. Integral equation (5) was solved directly both for positive and negative streamers. To reduce the computation time, we calculated the VUV photon flux using a coarse mesh and averaging photon generation and absorption over 10 point regions.

### Modeling results

Figure 3 compares the spatial distributions of the main parameters near the streamer head for positive and negative polarities. These distributions correspond to the instants of time when the heads of the positive and negative streamers are at the same distance from the high-voltage electrode. The negative streamer is slower; therefore, we have $t = 9.1$ ns (relative to the instant of applying the voltage) for the negative streamer and $t = 7.0$ ns for the positive streamer. It follows from the axial profiles plotted in figure 3(a) that the positive streamer has not only a higher velocity, but a higher peak electric field in the ionization wave as well.

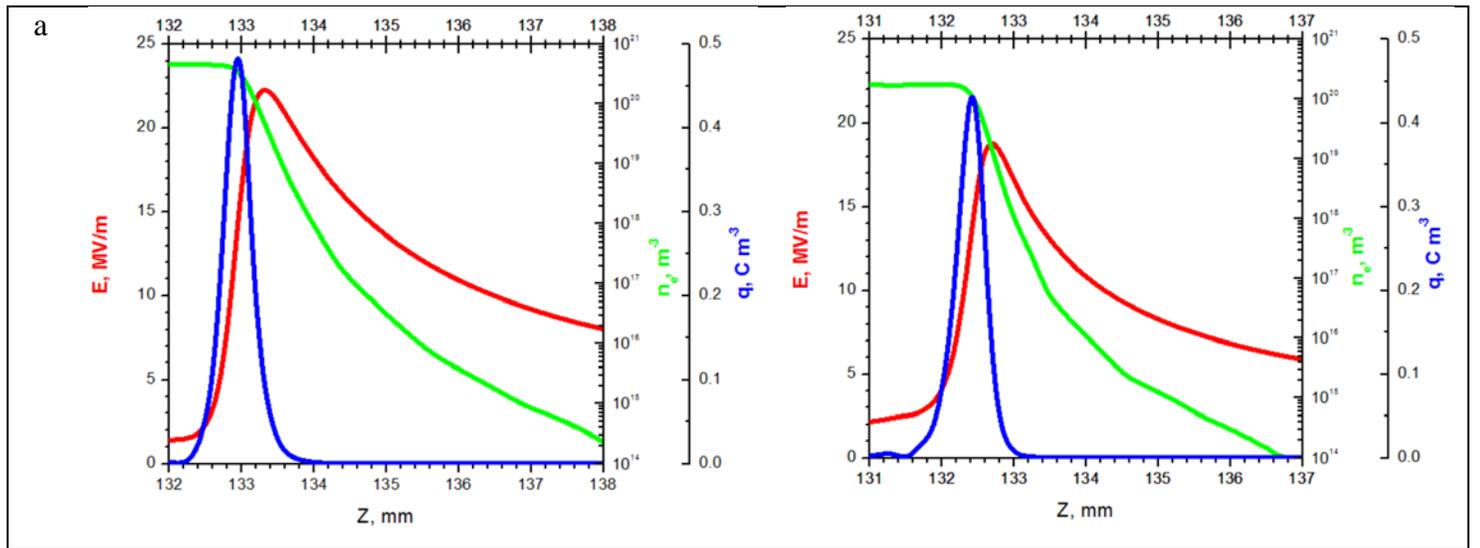

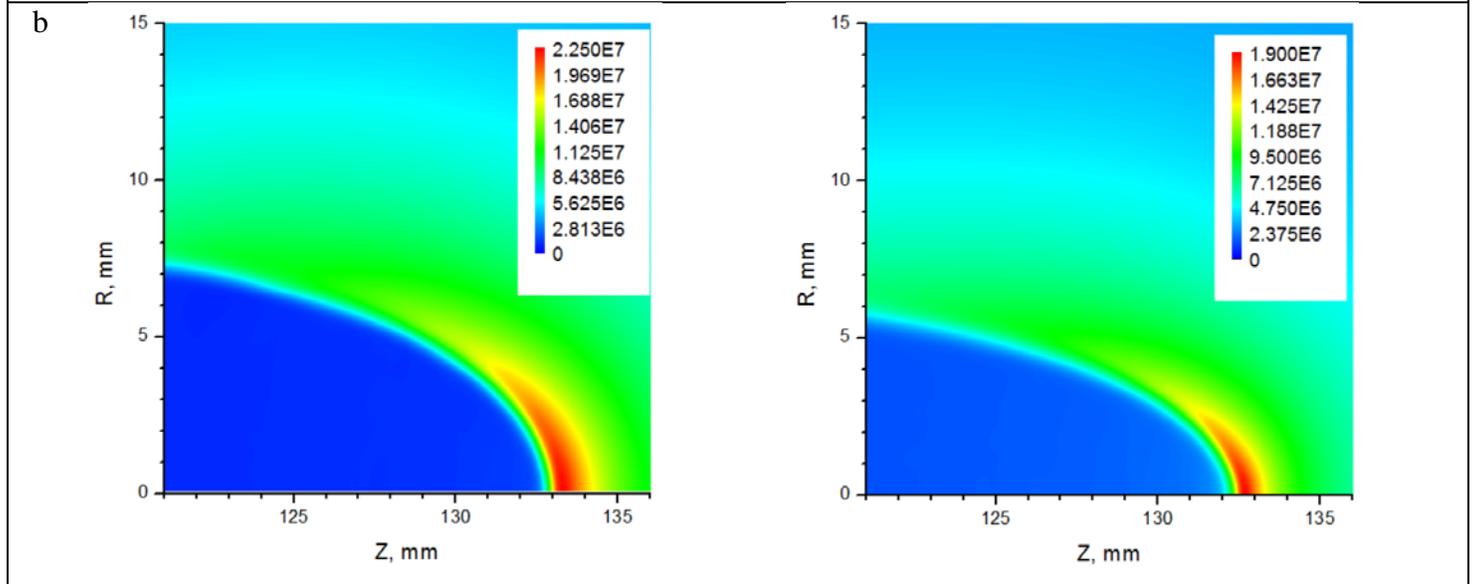

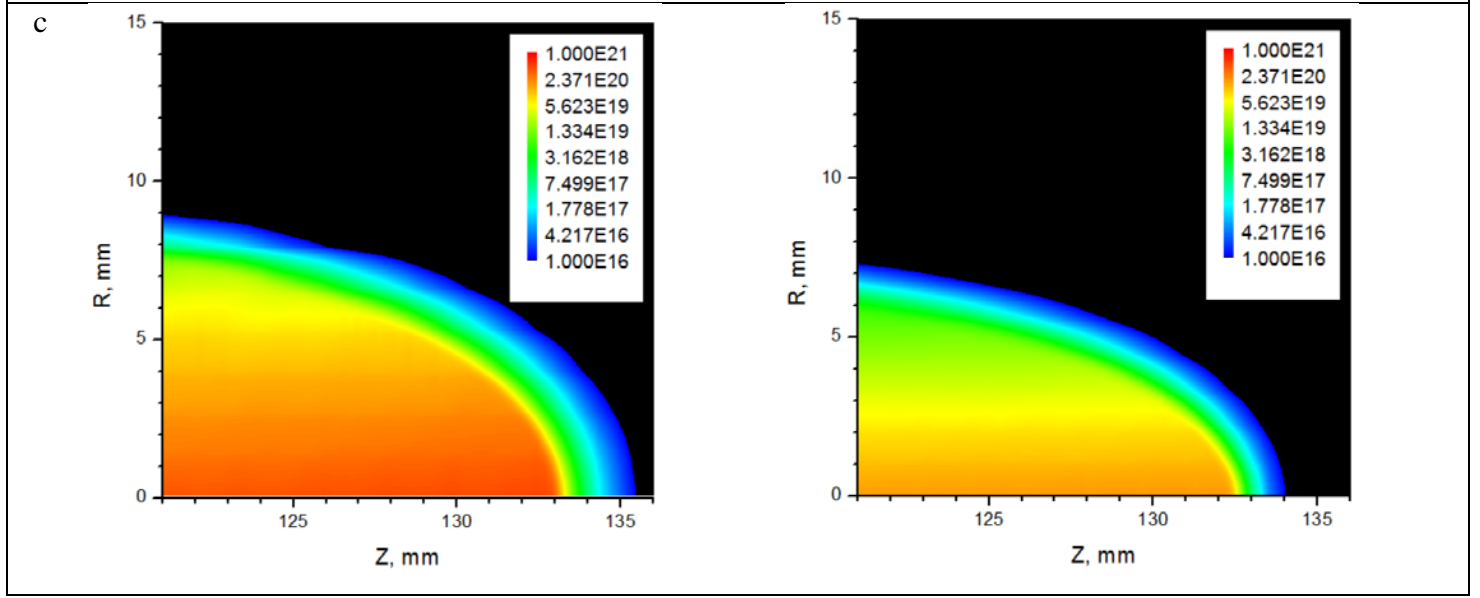

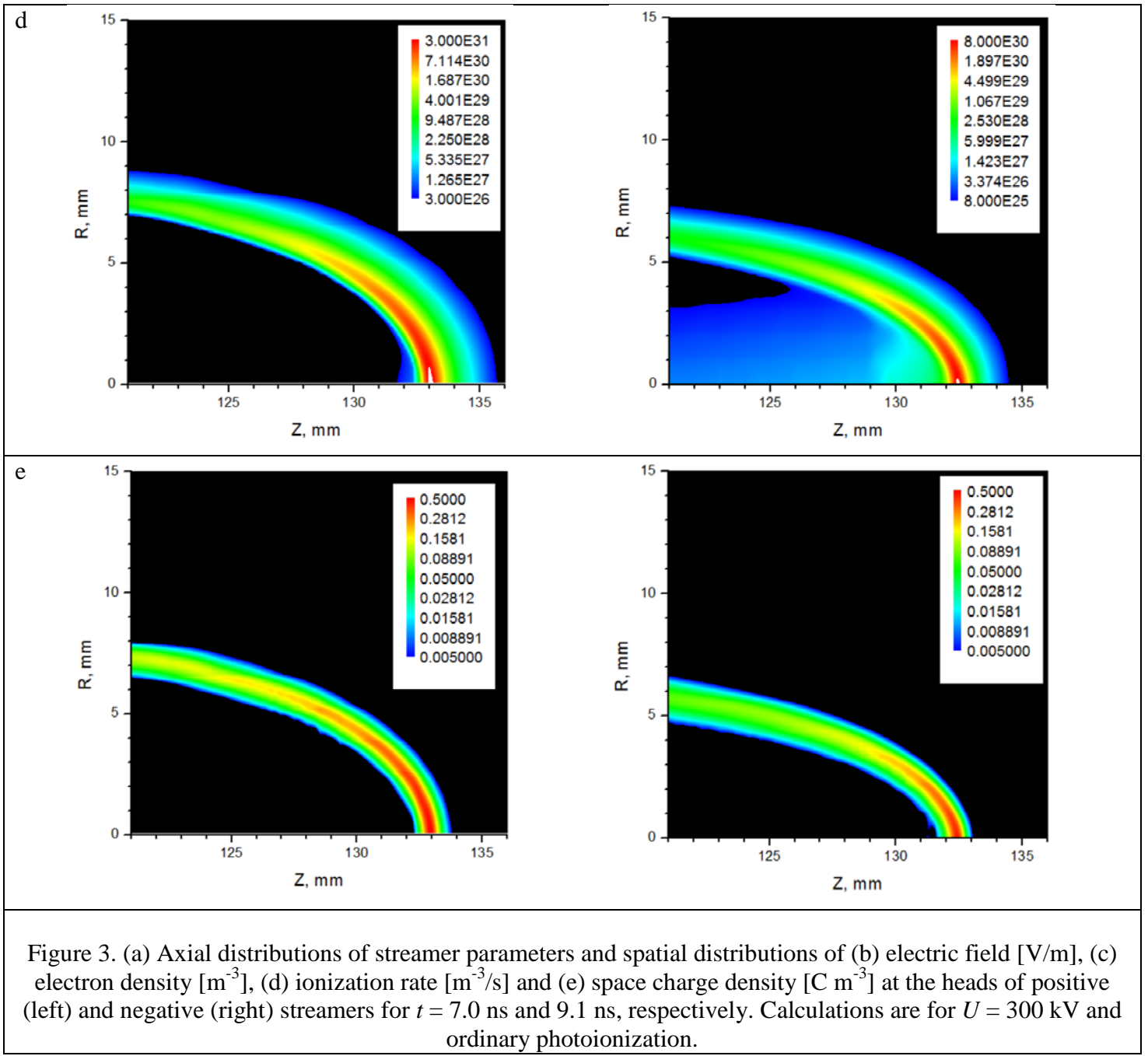

Figure 3. (a) Axial distributions of streamer parameters and spatial distributions of (b) electric field [V/m], (c) electron density [m$^{-3}$], (d) ionization rate [m$^{-3}$/s] and (e) space charge density [C m$^{-3}$] at the heads of positive (left) and negative (right) streamers for $t = 7.0$ ns and 9.1 ns, respectively. Calculations are for $U = 300$ kV and ordinary photoionization.

The peak electric field reaches 24 MV/m for positive polarity and only 21.5 MV/m for negative polarity. The differences are also obvious for the space charge density in the streamer head and the electron density in the streamer cannel. We have space charge densities of 0.49 and 0.42 C m$^{-3}$ for positive and negative polarities, respectively. The electron density is $5 \times 10^{20}$ m$^{-3}$ in the channel of the positive streamer and $1.4 \times 10^{20}$ m$^{-3}$ in the channel of the negative streamer. The electric field in the streamer channel just after its head is 1.2 MV/m for positive polarity and 2.2 MV/m for negative polarity.

Figure 3(b) shows the spatial distributions of electric field in the head of the positive and negative streamers. Here, the region of high electric field is larger for positive polarity. As a result, the transverse radii of the regions with high ionization rate (figure 3(d)) and with high electron density (figure 3(c)) are also higher for the positive streamer. The radius of the channel in which the density of the radiating $N_2$ molecules (primarily the $N_2(C)$ state) excited by high-energy electrons in the streamer head should be larger for positive polarity, too. Therefore, positive streamers registered in experiments should be thicker than negative streamers, in agreement with optical observations [8]. The region of space charge (figure 3(e)) is close to the region of high electric field because the transition from a low electric field in the channel to a high electric field in the streamer head is due to this space charge.

Figure 3 demonstrates that the positive and negative streamers have qualitatively similar behavior. However, the quantitative streamer parameters depend on voltage polarity. Hereafter the differences between the characteristics of the positive and negative streamers and reasons why these differences exist are analyzed in detail.

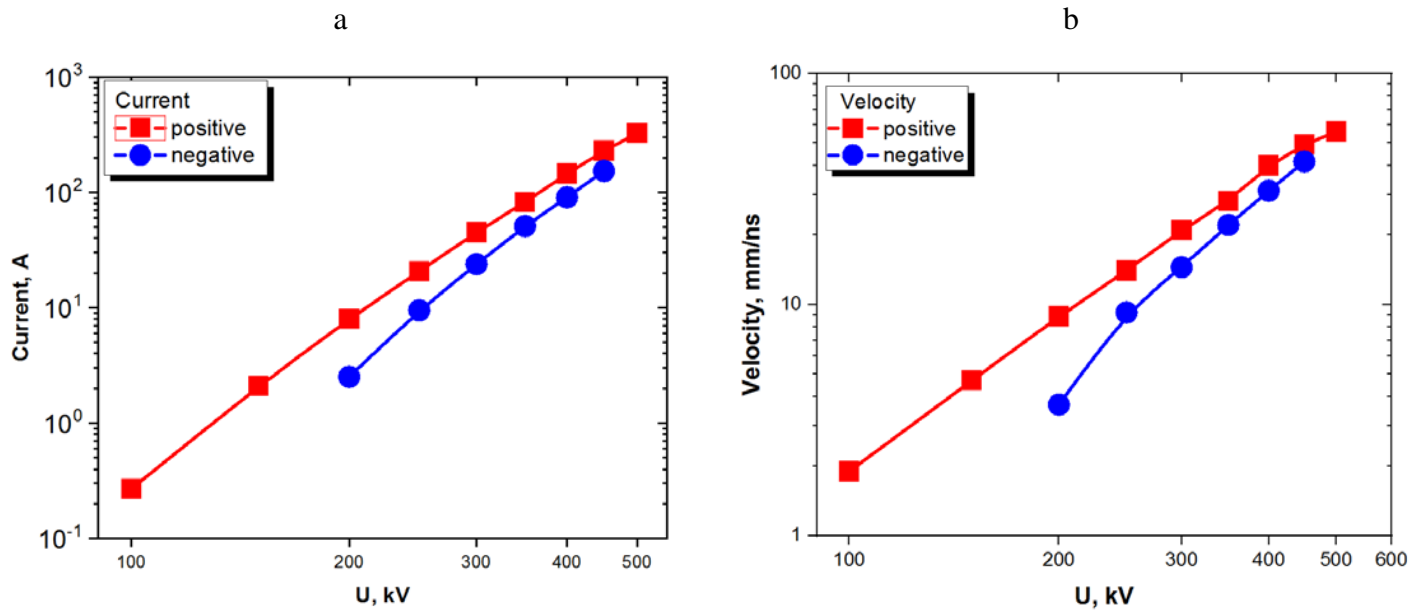

Figure 4. (a) Discharge current and (b) streamer velocity for various polarities versus applied voltage. Calculations are for a distance of $Z_{head} \approx 80$ mm between the streamer head and the high-voltage electrode.

Figure 4 demonstrates the calculated velocity and discharge current for the positive and negative streamers versus the applied voltage. The voltage is varied from the minimum values (75 and 150 kV for positive and negative polarities, respectively) to the values corresponding to breakdown in a uniform electric field (500 and 450 kV for positive and negative polarities, respectively). The average electric field in the 14 cm air gap is varied in the range 5.3 – 35.7 kV/cm for positive polarity and in the range 10.6 – 32.1 kV/cm for negative polarity. Thus we cover the total voltage range in which streamers can be initiated. It follows from figure 4 that, for both polarities, the streamer velocity increases with increasing applied voltage much smaller than does the discharge current. The streamer current increases almost by three orders of magnitude (from 0.8 to 300 A) as the voltage grows from the minimum value to the maximum one. In this case, the streamer velocity calculated in the middle of the discharge gap increases only by an order of magnitude, from $3\times10^6$ to $6\times10^7$ m/s. This difference in the voltage dependencies of the streamer velocity $v_S$ and current $i$ follows from the relationship $i \sim \tau_S v_S$ [1], where $\tau_S$ is the linear (per unit length) charge density. This quantity also increases with increasing applied voltage $U$ as $\tau_S \approx CU$, where $C$ is the linear capacitance of the streamer. The value of $C$ slightly (logarithmically) depends on streamer dimensions and its variation with increasing voltage is much smaller than variations of $i$ and $v_S$. Therefore, the ratio between $i$ and $v_S$ increases almost linearly with $U$, in agreement with the results presented in figure 4.

Figure 5 shows the reduced electric field at the streamer head and in the channel versus the applied voltage. The peak reduced electric field at the head of the positive streamer grows from 600 Td for $U = 100$ kV to 1050 Td for $U = 500$ kV. The peak values of the electric field for the negative streamer are close to those for the positive streamer; at negative polarity we have 600 Td for $U = -200$ kV and 900 Td for $U = -450$ kV. The electric field at the head of the negative streamer is somewhat lower than that at the head of the positive streamer, whereas there is an opposite relation for the electric field in the streamer channel, $E_{ch}$. Its value is determined as the minimum electric field in the channel to avoid the high electric field regions near the high-voltage electrode and at the head. The values of $E_{ch}$ for the negative streamer are approximately twice those in the channel of the positive streamer. The reduced electric field $E_{ch}/n$ ($n$ is the gas number density) grows with increasing voltage from 20 to 42 Td in the channel of the positive streamer and from 50 to 63 Td in the channel of the negative streamer.

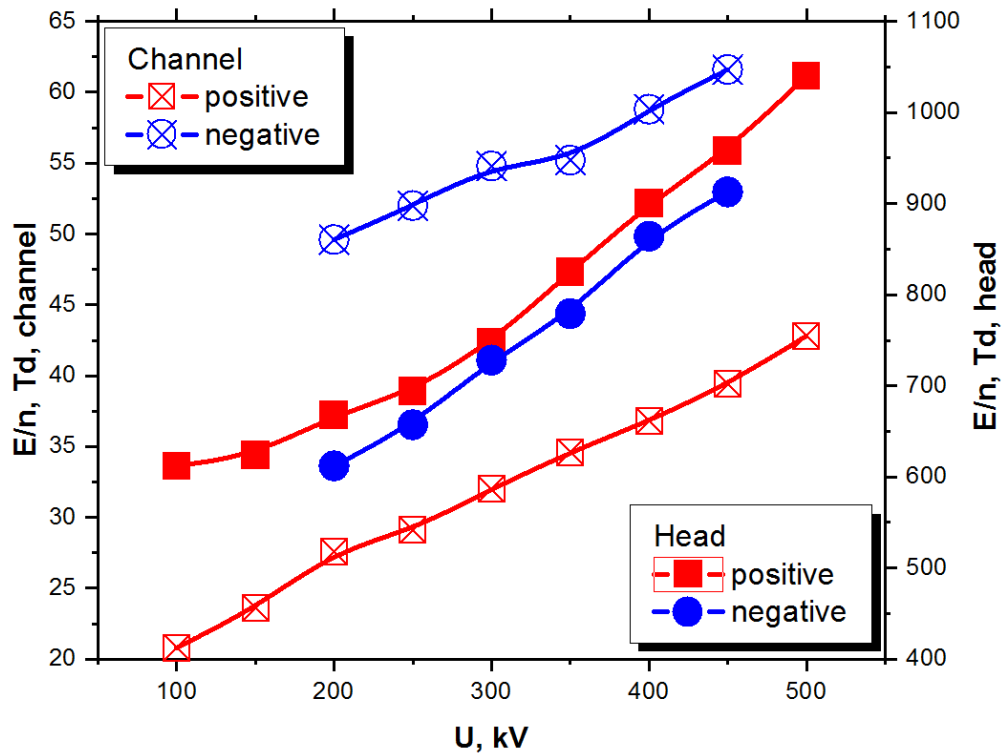

Figure 5. Reduced electric field for various polarities as a function of applied voltage. Electric fields are given at the streamer heads and in the channels. Calculations are for a distance of $Z_{head} \approx 80$ mm between the streamer head and the high-voltage electrode.

An important characteristic of a long streamer is the electric field in its channel after bridging the gap, $E_b$. This field is also referred to as the ambient electric field required for sustained streamer propagation [22]. The values of $E_b$ allow determining the limiting streamer length that is approximately the applied voltage divided by $E_b$. This field measured for positive streamers in long air gaps under standard conditions is in the range 4.5-5.0 kV/cm [1]. For negative streamers, the field $E_b$ was estimated when considering the streamer zone of the negative leader progressing in long (1 – 10 m) air gaps. The estimated values of $E_b$ are in the range 10-16 kV/cm for negative polarity [1]. We determined the field $E_b$ using the calculated data. The calculated voltage required for bridging the 14 cm gap was 75 kV for positive polarity and 150 kV for the negative polarity streamer. The corresponding values of the electric field required for bridging the gap were 5.3 kV/cm for the positive streamer and 10.1 kV/cm for the negative streamer, in agreement with available experiments.

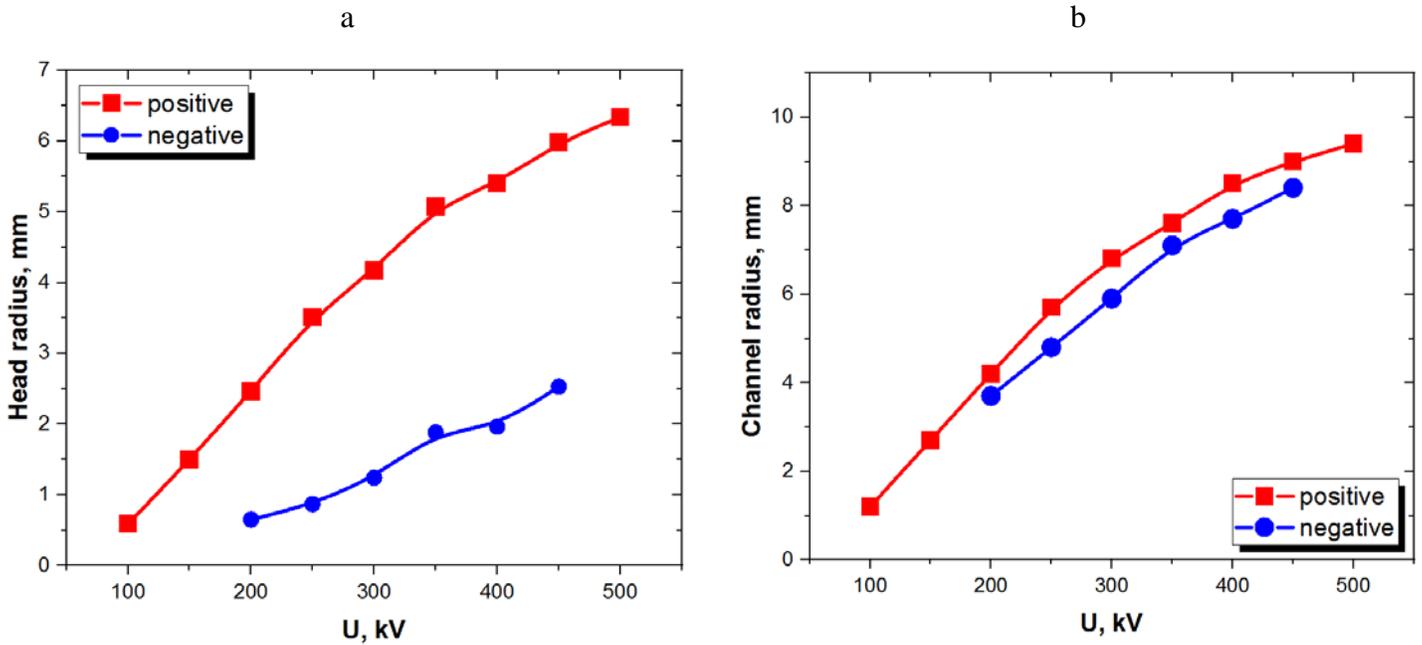

Figure 6. (a) The curvature radius of the head and (b) the channel radius corresponding to the peak radial electric field as a function of applied voltage. Calculations are for a distance of $Z_{head} \approx 80$ mm between the streamer head and the high-voltage electrode.

Based on the results obtained, it is possible to explain why the average electric field $E_{ch}$ in the channel of the negative streamer is much higher than that in the channel of the positive streamer. The value of $E_{ch}$ is controlled by the plasma conductivity in the channel and by the total streamer current, which charges the "capacitor" formed by the channel during the discharge development: $E_{ch} \sim i/(\sigma \pi r_{ch}^2)$, where $\sigma$ is the plasma conductivity in the channel and $r_{ch}$ is the effective radius of the current-carrying channel. The conductivity is controlled by the electron density $n_e$ in the channel. The radius $r_{ch}$ can be estimated using the radial profile of $n_e$. It follows from the calculations that the values of $i$, $n_e$ and $r_{ch}$ are higher for the positive streamer (figures 3 and 4). However, the ratio $i/n_e$ is approximately the same for both polarities, whereas the value of $r_{ch}^2$ for the positive streamer is approximately twice that for the negative streamer. It may be concluded that the higher electric field in the channel (and the higher value of $E_b$) of the negative streamer follows from the shorter effective radius of its current-carrying channel. The difference in the values of $r_{ch}$ for the negative and positive streamers is explained by the higher electric field in the streamer head for positive polarity (figures 3 and 5).

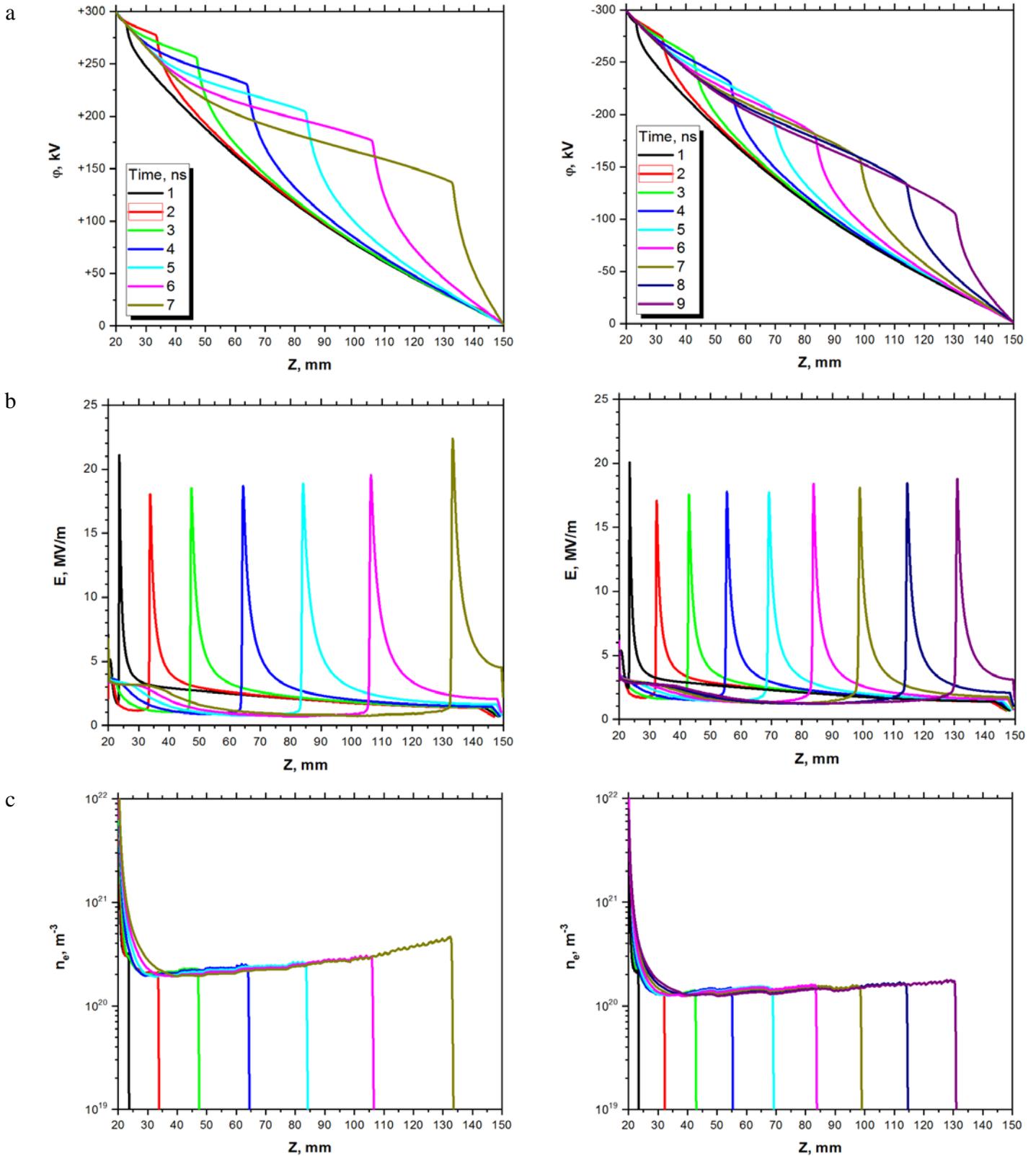

Figure 7. Axial profiles of (a) electrostatic potential, (b) electric field and (c) electron density in the channels of (left) positive and (right) negative streamers for $U = 300$ kV. The curves are plotted at equal time steps of 1 ns.

The radius of the streamer head increases with rising applied voltage for both polarities due to more efficient ionization in this region (figure 5(a)). The radius of the streamer channel can also be determined using the peak radial electric field at the side surface of the channel. This radius is almost independent of voltage polarity and increases somewhat slower with applied voltage than does the effective radius of the current-carrying channel. For the voltages being close to the thresholds of streamer initiation, the radius corresponding to the peak radial electric field is approximately 1 mm for positive polarity and around 3 mm for negative polarity. An increase in the applied voltage leads to the faster development of a radial ionizing wave. As a result, the radius of the channel reaches 9 mm for the positive streamer and 7.5 mm for the negative streamer when the applied voltage reaches 500 kV (figure 6(b)).

The properties of a streamer change during its development because (i) the increasing voltage drop across the streamer channel leads to a decrease in the potential of the streamer head, (ii) the local electric field increases when the streamer head approaches the opposite electrode, and (iii) the amount of seed electrons produced by photoionization is accumulated in the discharge gap. The third effect is of particular importance when a streamer propagates through the media with density discontinuities [23-25].

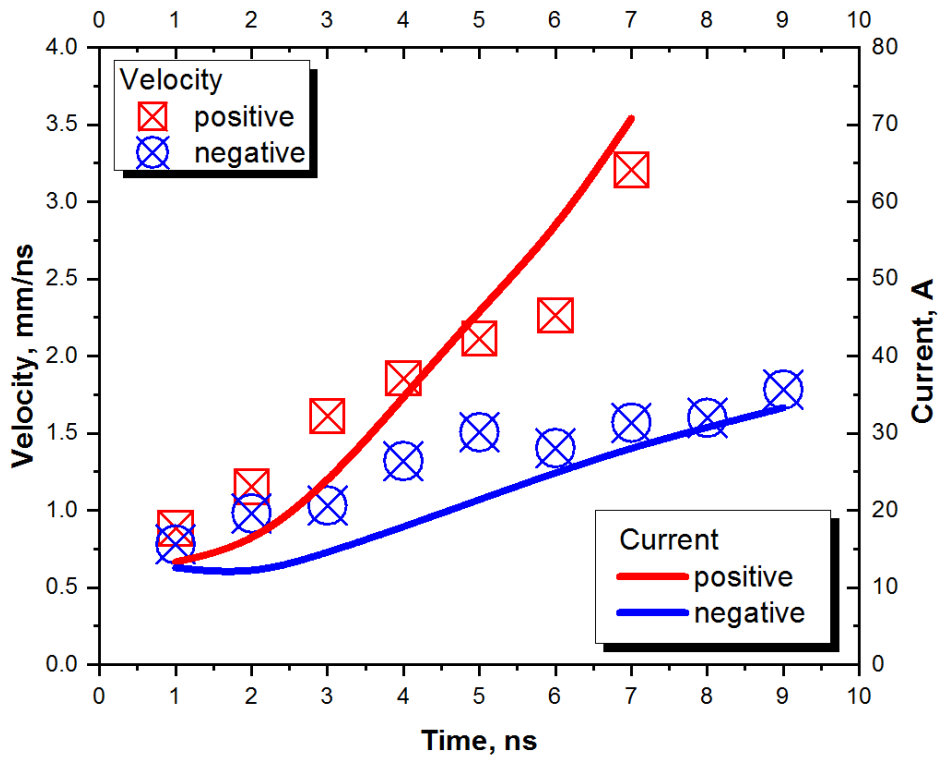

Figure 8. The temporal evolution of the streamer velocity and discharge current for positive and negative streamers at $U = 300$ kV.

Figures 7-9 demonstrate the temporal evolution of the main discharge parameters during streamer propagation in the gap for an applied voltage of $U = 300$ kV. Figure 7(a) shows the axial profiles of the electrostatic potential at various instants. The voltage drop increases up to ≈ 150 kV for positive polarity and up to ≈ 200 kV

for negative polarity as the streamer heads move away from the high-voltage electrode. The peak electric field at the streamer head and the average electric field in the channel reach the corresponding quasi-stationary values (figure 7(b)). According to the axial profiles of the electron density (figure 7(c)), electrons and ions have no time to recombine in the channel for this period of time. The electron density in the channel of the positive streamer increases as the streamer head approaches the opposite electrode. The average electron density in the channel for positive polarity is 50-100% higher than that in the channel for negative polarity. A higher voltage drop across the channel of the negative streamer leads to a higher decrease in the potential of the streamer head. As a result, even an increase in the local electric field near the opposite electrode does not cause a noticeable increase in the electron density in the negative streamer.

Figure 8 demonstrates the temporal evolution of the streamer velocity and current during the discharge development. In the beginning, the streamer velocity is low since the calculations take into account a finite (1 ns) time of voltage rise at the high-voltage electrode. The velocities of the positive and negative streamers increase up to 1 mm/ns after the first nanosecond. Further discharge development depends on voltage polarity. The velocity of the positive streamer rises in time and reaches 2-2.25 mm/ns by the fifth nanosecond. The velocity of the negative streamer shows saturation and is ~1.5 mm/ns for $t = 4$-8 ns due to a higher voltage drop across the streamer channel. The streamer current increases during the discharge development since the current is proportional to the streamer velocity: $i \sim \tau_s v_s$ [1]. The streamers accelerate when approaching the grounded electrode. At this instant the discharge current reaches 35 A for negative polarity and 65 A for positive polarity.

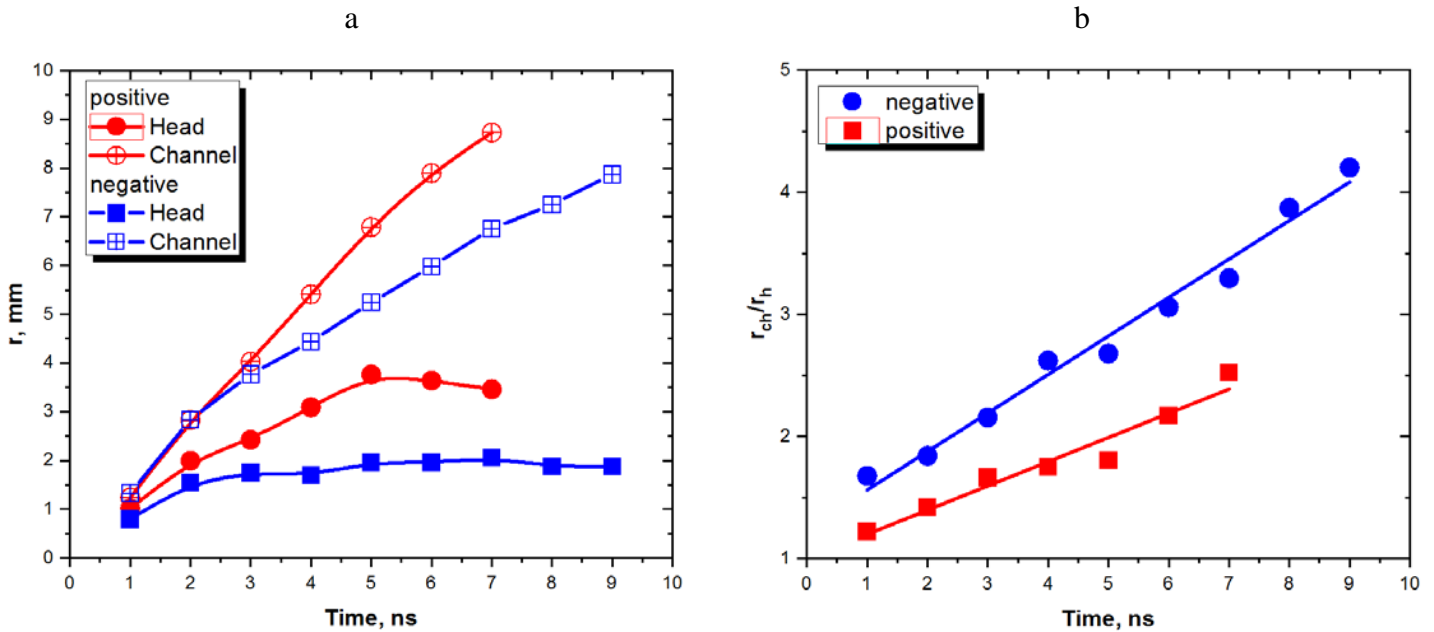

Figure 9. (a) The temporal evolution of channel and head radii. (b) The ratio of channel radius $r_{ch}$ to head radius $r_h$. Calculations are for positive and negative streamers at $U = 300$ kV.

The maximum channel radius $r_{ch}$ corresponding to the peak radial electric field also increases with the discharge development due to a radial ionizing wave. The electric field on the side surface of the channel is much lower than that at the streamer head. Therefore, the streamer radius remains much lower than the streamer length $L$. Figure 9 demonstrates the temporal evolution of the channel radius $r_{ch}$ and the curvature radius of the streamer head $r_h$ for both polarities. The channels of the positive and negative streamers expand with constant radial velocities. The ratio between the channel length and radius is almost independent of time and equals $L/r_{ch} \approx 10$ for positive polarity and $L/r_{ch} \approx 14$ for negative polarity.

The curvature radius corresponding to the peak electric field at the streamer head also increases during streamer development (figure 9). This radius for the positive streamer is almost twice that for the negative streamer. At the same time the electric field at the head of the positive streamer is higher than that of the negative streamer (figure 5). This difference is associated with a larger voltage drop across the channel of the negative streamer and with a more efficient expansion of this channel just behind the streamer head. Due to this expansion, the ratio of the channel radius to the head radius is higher for the negative streamer (figure 9(b)). The spatial distribution of electric field in the vicinity of the positive streamer head is close to the distribution near the hemispherical tip of a cylinder, whereas the distribution near the negative streamer head is close to the distribution near a cone with a rounded vertex. A similar difference between the electric field spatial distributions were numerically obtained at the heads of positive and negative streamers in short (1 cm) air gaps [12]. In this simulation, as the streamer advanced, it became thicker and faster, also in agreement with our calculations.

It may be concluded that the differences between the positive and negative streamers in long air gaps agree with observations [1, 8] and with previous simulations of streamer discharges in short air gaps [15]. In particular, the average electric field in the streamer channel is higher for negative polarity, whereas the peak electric field in the streamer head, velocity, electron density in the channel, discharge current, head radius and channel radius are higher for positive polarity.

In the following, we analyze the physical mechanisms leading to the differences between the positive and negative streamers.

## The effect of photoionization on streamer properties

In the system of equations (1) – (5), the only parameter that depends on the polarity of applied voltage is the direction of electron drift. All other processes described by equations (1) – (5) are independent of voltage polarity. Under the conditions studied, the electron energy distribution is controlled by the local electric field. Therefore, in the local approximation used in this work the ionization rate, absolute value of the electron drift velocity and the rates of photoionization and electron-ion recombination are also independent of the direction of the local electric field.

To clarify the role of these processes in streamer development, let us consider the dominant processes occurring on the streamer axis near the point at which the electric field peaks in the streamer head. In the system of coordinates moving with the streamer velocity $v_s$, it follows from a quasi-stationary spatial distribution of streamer characteristics that the electron density $n_e$ and other streamer characteristics do not change with time. In this case, when neglecting 2D effects, we have the electron balance equation

$$\frac{dn_e}{dt} = k_i n n_e - (v_S \pm v_d)\frac{dn_e}{dx} - n_e \frac{d(v_S \pm v_d)}{dx} = 0 \ . \tag{6}$$

Here, the first term on the right-hand side of equation (6) describes electron-impact ionization of neutral particles with the gas number density $n$, $k_i$ is the ionization coefficient, and the second and third terms correspond to the axial electron transport due to the gradients of $n_e$ and electron drift velocity $v_d$, respectively. In (6), we have $+v_d$ for negative streamer and $-v_d$ for positive streamer. At the point corresponding to the peak electric field, the third term in (6) vanishes. Then, equation (6) is reduced to a simple relation between the local ionization rate and the term describing the electron transport with the velocity $v_S \pm v_d$ relative to the ionization front:

$$k_i n = (v_S \pm v_d)\frac{dn_e}{dx}\frac{1}{n_e} \tag{7}$$

In (7), the voltage polarity affects the sign of the electron drift velocity only.

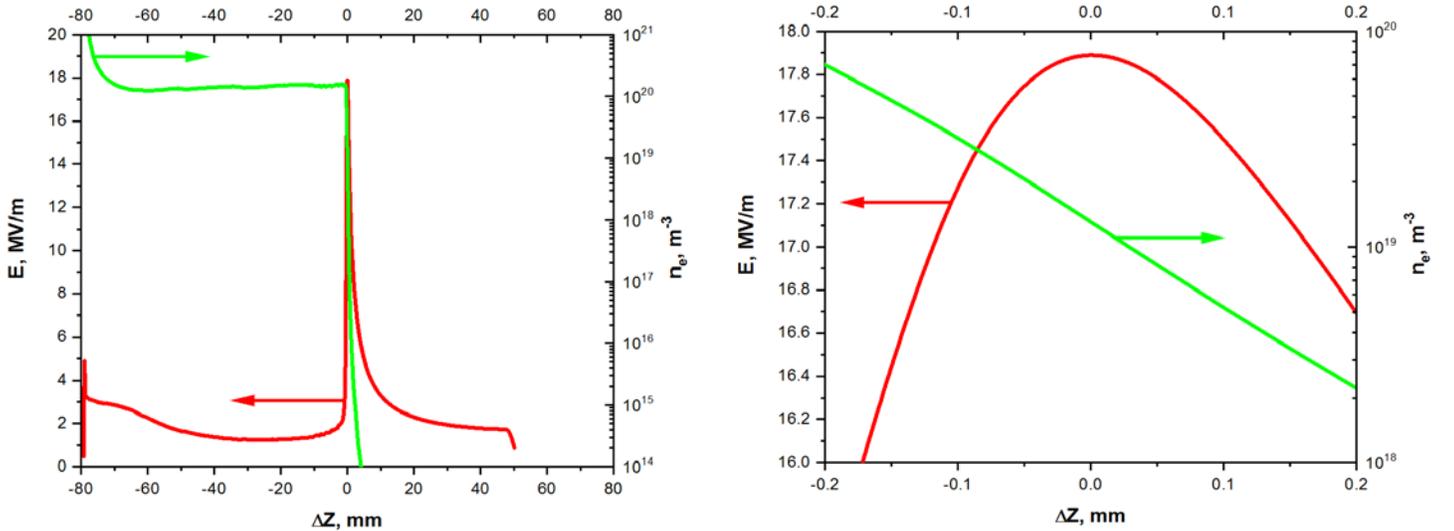

Figure 10. The axial distributions of electric field and electron density at the head of the negative streamer for $U = -300$ kV and $t = 7.0$ ns. Left and right figures correspond to different space scales. Calculations are for a distance of $Z_{head} \approx 100$ mm between the streamer head and the high-voltage electrode.

Using (7), we analyze the relative contributions of various fluxes to the electron balance at the ionization front of the streamer discharge. Firstly, we consider the results of the 2D numerical simulation for the negative streamer. Figure 10 shows the axial profiles of electric field and electron density at the head of the negative streamer in atmospheric pressure air for $U = -300$ kV when the average reduced electric field in the gap is 80 Td. The curves correspond to the time $t = 7$ ns when the position of the streamer head is $Z_{head} \approx 100$ mm. The peak electric field and the electron density at this point are 17.9 MV/m and $1.3\times10^{19}$ m$^{-3}$, respectively. Here, the electron drift velocity is $v_d = 4.8 \times 10^5$ m/s and the streamer velocity is $v_S = 1.56 \times 10^7$ m/s. Using these values, we have $k_i n = 1.2 \times 10^{11}$ s$^{-1}$ and $(v_S + v_d)\frac{dn_e}{dx}\frac{1}{n_e} = 1.4 \times 10^{11}$ s$^{-1}$. The difference between these values corresponding to the left-hand side and right-hand side of (7) is small and explained by neglecting 2D effects in (6) and (7) and by a slow variation of streamer characteristics with time.

Figure 11 demonstrates the axial profiles of electric field and electron density obtained from the 2D simulation of the positive streamer developing under the conditions corresponding to those for the results presented in figure 10. Here, the streamer properties correspond to the instant $t = 5.5$ ns when the streamer head is at $Z_{head} \approx 100$ mm. In this case, the peak electric field at the streamer head is 19.3 MV/m and the electron density at this point is $3 \times 10^{19}$ m$^{-3}$. In this case, the electron drift velocity and streamer velocity are $v_d = 5.1 \times 10^5$ m/s and $v_S = 1.92 \times 10^7$ m/s, respectively. Then, we have $k_i N = 1.46 \times 10^{11}$ s$^{-1}$ and $(v_S - v_d)\frac{dn_e}{dx}\frac{1}{n_e} = 1.3 \times 10^{11}$ s$^{-1}$. The difference between these values for the positive streamer also is small. At the point of peak electric field, the contribution of electron-impact ionization to the electron balance is just larger than the contribution of electron transport for the positive streamer and just smaller than the transport contribution for the negative streamer.

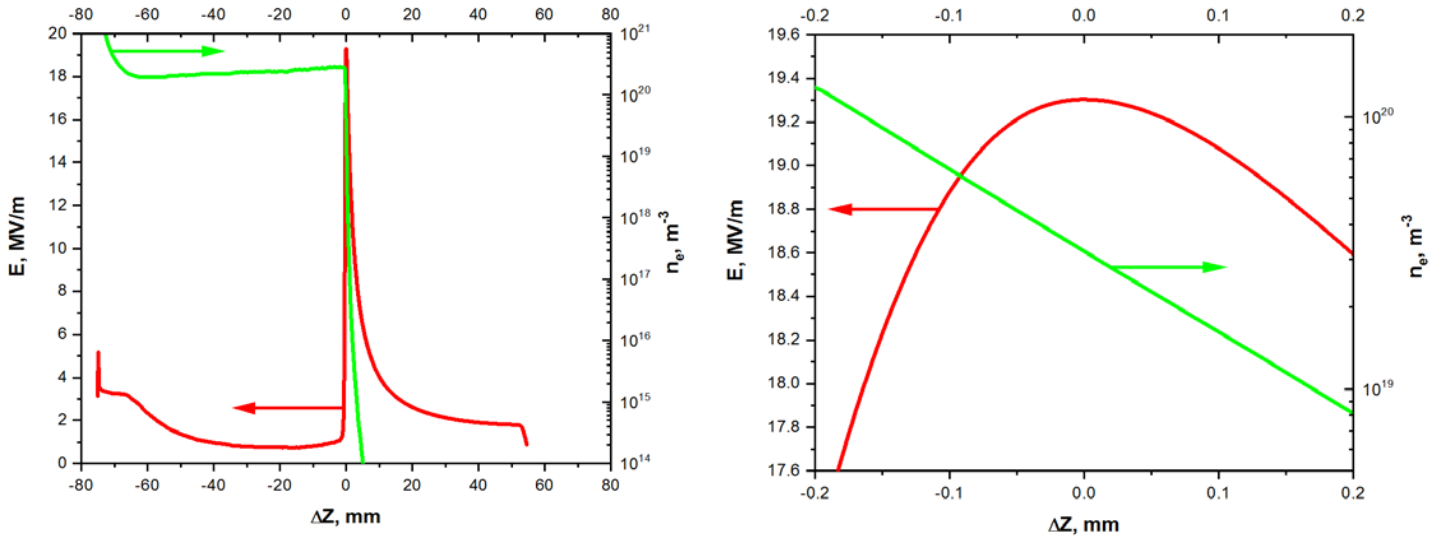

Figure 11. The axial distributions of electric field and electron density at the head of the positive streamer for $U = +300$ kV and $t = 5.5$ ns. Left and right figures correspond to different space scales. Calculations are for a distance of $Z_{head} \approx 100$ mm between the streamer head and the high-voltage electrode.

We consider strong streamers when the electron balance at the streamer head is governed by electron-impact ionization in the local electric field and by the propagation of the ionizing front. In this case, the contribution of electron drift to the electron balance is small because the electron drift velocity at the streamer head is much lower than the streamer velocity. Different directions of electron drift cause a small disturbance in the electron balance at the streamer head, only. However, the presence of seed electrons in front of the steamer head is required to form avalanches for both the positive and negative discharges. For this reason, photoionization must play an important role for both polarities.

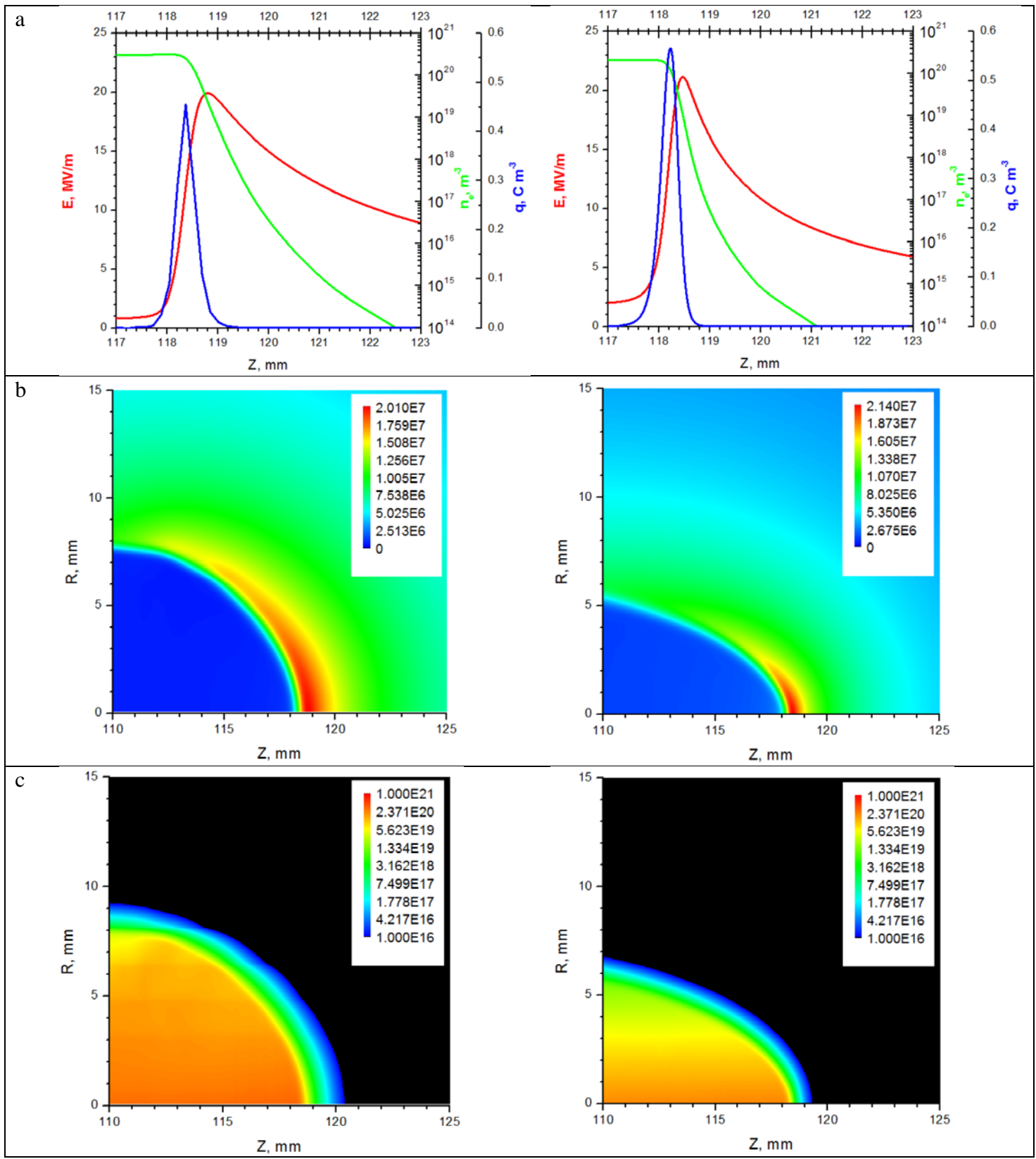

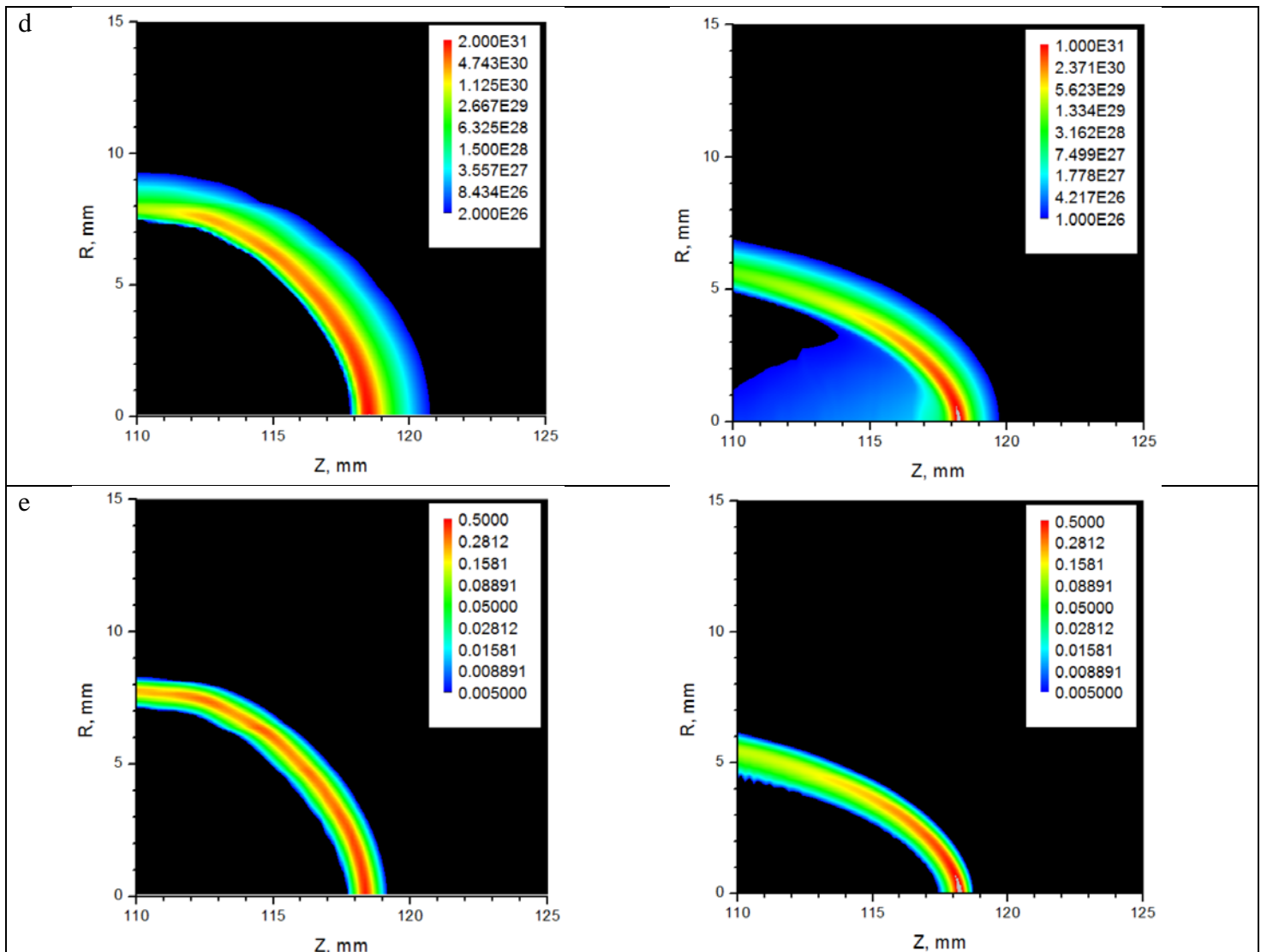

Figure 12. (a) Axial distributions of streamer parameters and spatial distributions of (b) electric field [V/m], (c) electron density [$m^{-3}$], (d) ionization rate [$m^{-3}$/s] and (e) space charge density [$C\,m^{-3}$] at the heads of positive (left) and negative (right) streamers for $t = 7.0$ ns and 8.2 ns, respectively. Calculations are for $U = 300$ kV and the photoionization rate decreased by a factor of 10.

It may be concluded from the analysis of the processes at the streamer ionization front that photoionization plays a key role for the development of both positive and negative discharges at high applied voltages. This conclusion is well known for positive streamers, whereas the importance of photoionization for negative streamers has not been adequately studied. We consider quantitatively the effect of photoionization on the properties of positive and negative streamers in long air gaps.

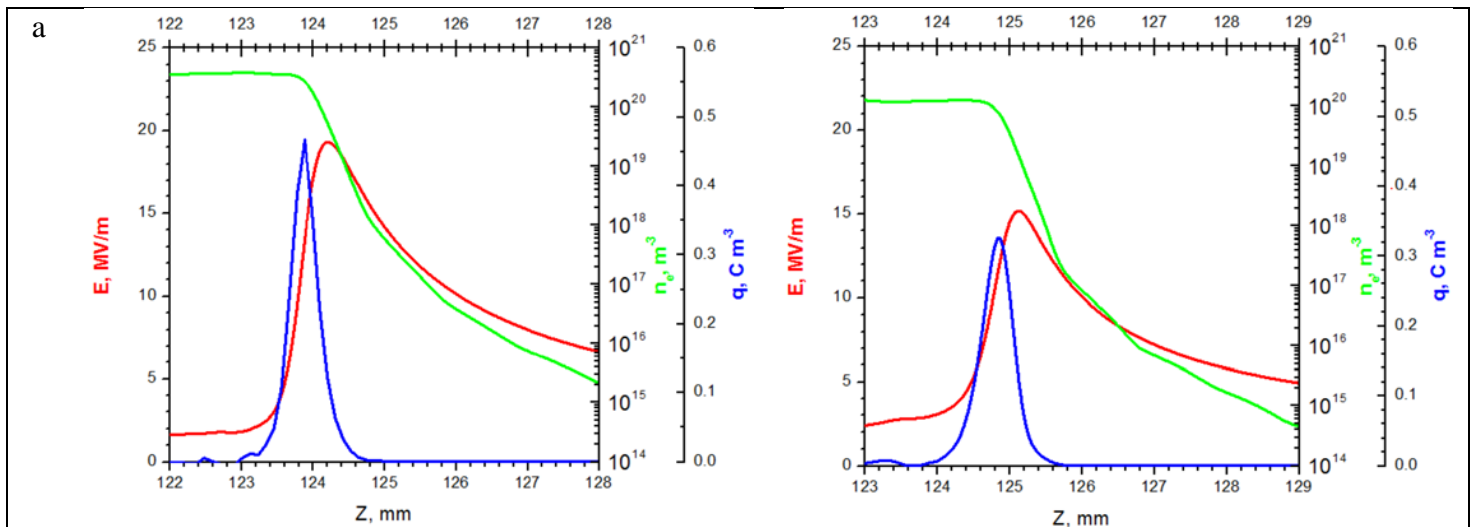
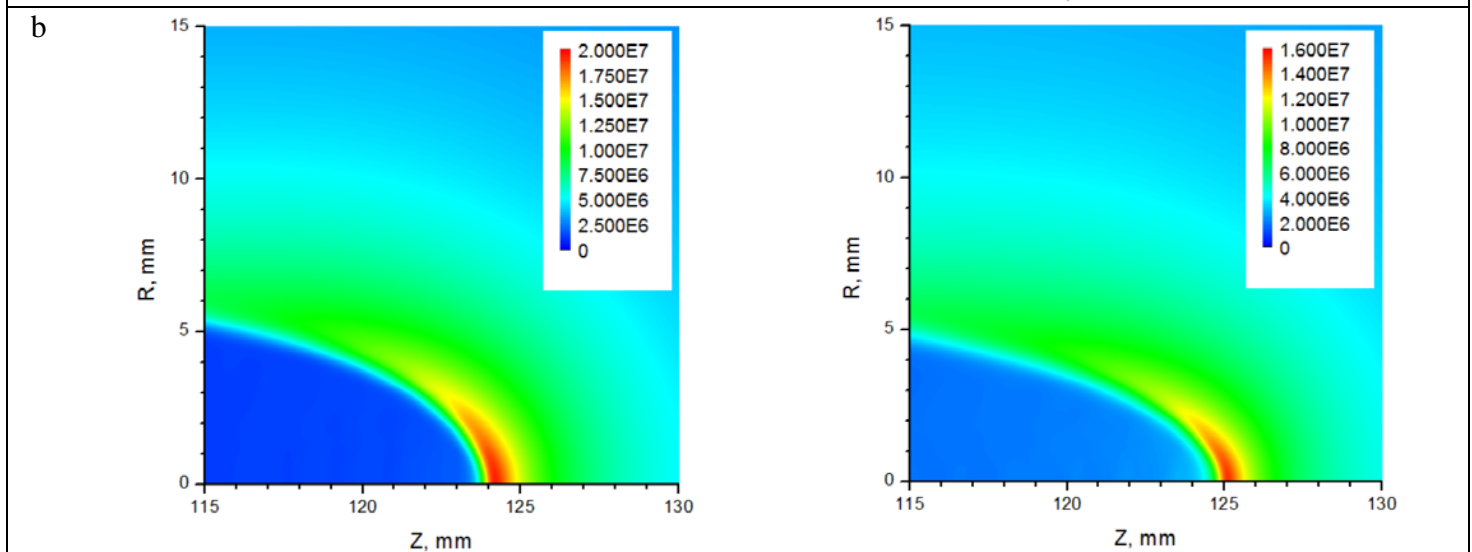
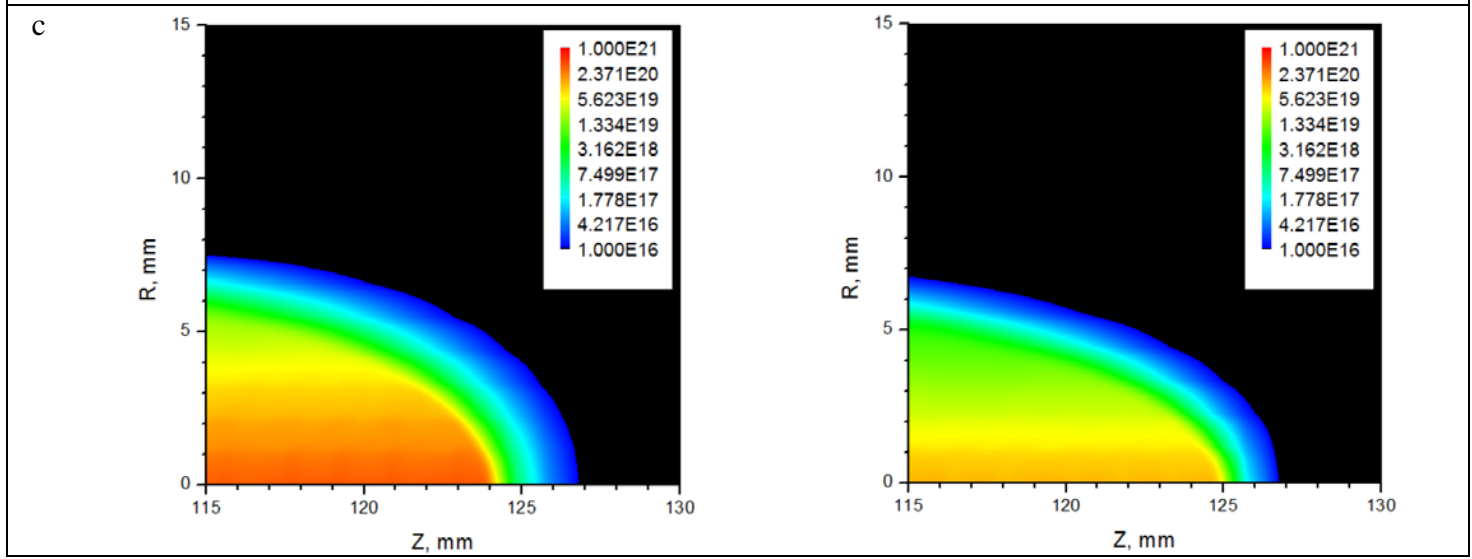

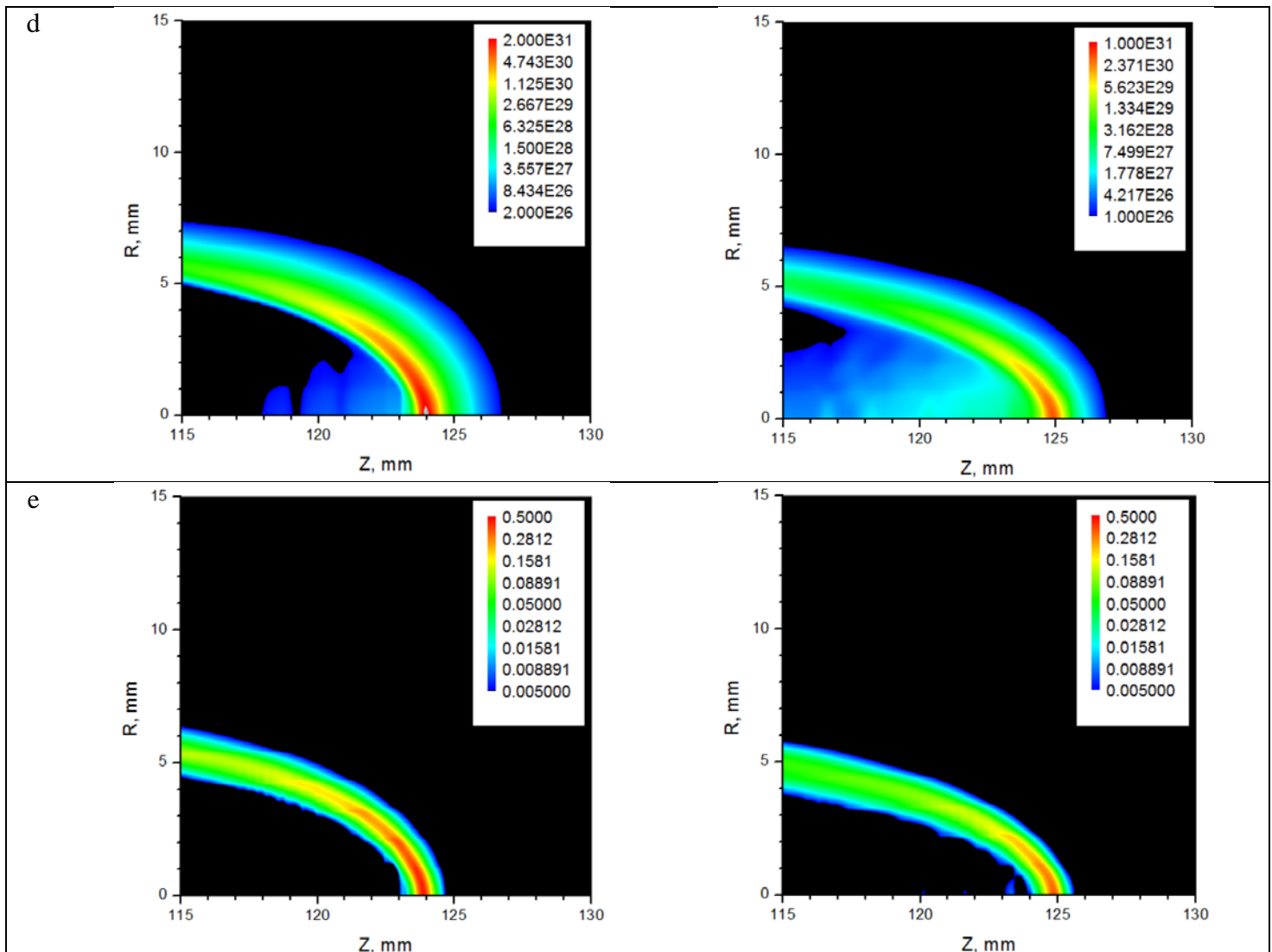

Figure 13. (a) Axial distributions of streamer parameters and spatial distributions of (b) electric field [V/m], (c) electron density [$m^{-3}$], (d) ionization rate [$m^{-3}$/s] and (e) space charge density [C $m^{-3}$] at the heads of positive (left) and negative (right) streamers for $t = 6.5$ ns and 9.1 ns, respectively. Calculations are for $U = 300$ kV and the photoionization rate increased by a factor of 10.

Figures 12 and 13 demonstrate the calculated results for the positive and negative streamers in atmospheric pressure air at an applied voltage of 300 kV. Calculations are made for various efficiencies of photoionization. Experimentally, photoionization efficiency can be increased by adding a gas with a large photoionization cross-section or decreased by adding a molecular gas for which the dominant process of photon absorption in the range 96-110 nm is photodissociation rather than photoionization. The results shown in figure 12 are obtained when decreasing the photoionization rate by a factor of 10 (absorbing additive), whereas the results shown in figure 13 correspond to the calculations when the photoionization rate is increased by a factor of 10 (easily ionized additive).

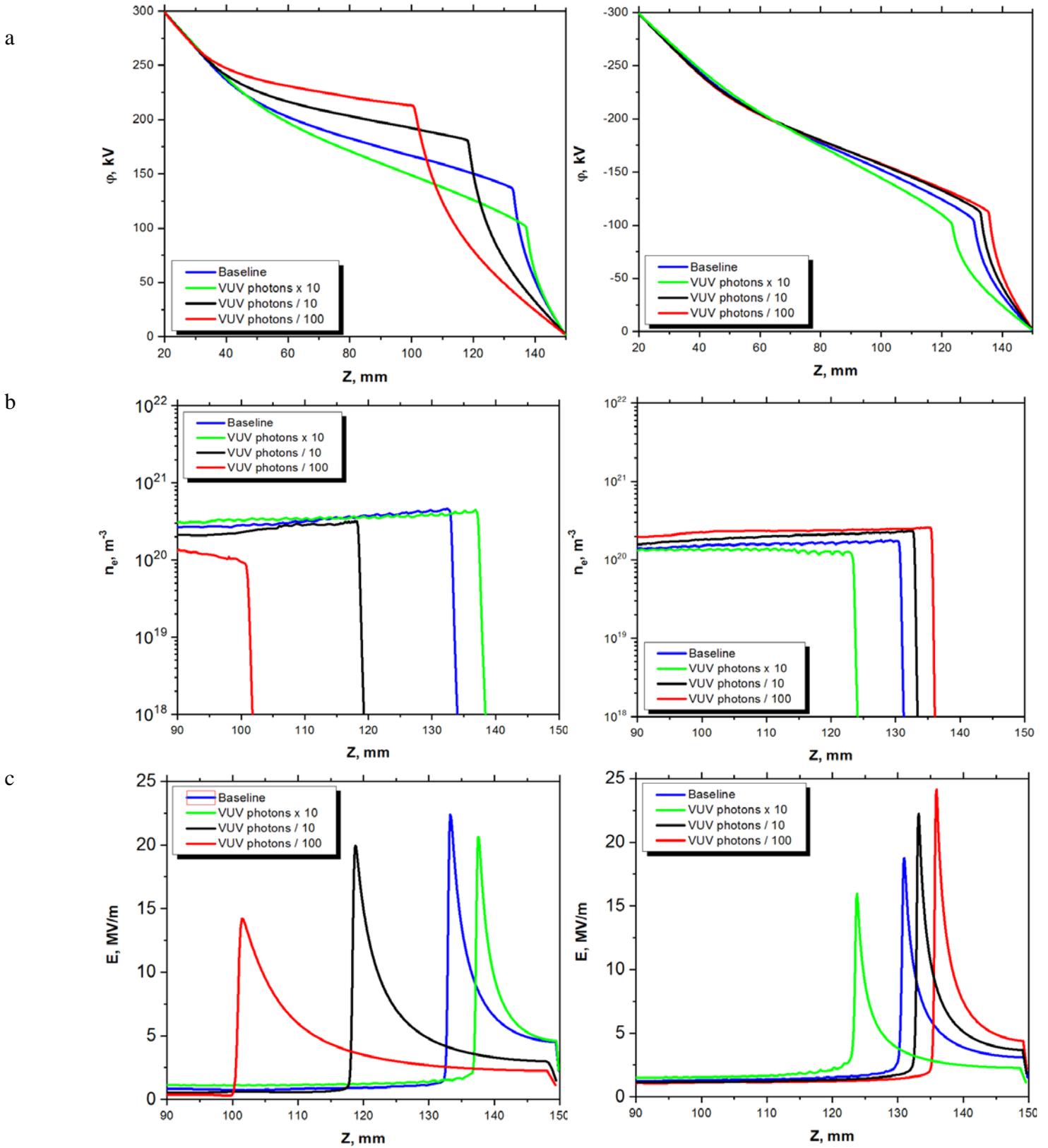

Figure 14. Axial profiles of (a) electrostatic potential, (b) electron density, and (c) electric field for various photoionization rates and discharge polarities. Curves correspond to (left) positive polarity for $t = 7$ ns and (right) negative polarity for $t = 9.1$ ns. The applied voltage is 300 kV.

It follows from these results that the variation in the efficiency of photoionization affects the positive and negative streamers in different ways. The decrease in the photoionization rate in front of the head of the positive streamer leads to a decrease in the streamer velocity, electron density in the channel and peak electric field at the head, whereas the head radius becomes higher.

The opposite effect is obtained for the negative streamer. Here, the decrease in the photoionization rate causes an increase in the streamer velocity, electron density in the channel and peak electric field at the head, whereas the radius of the head remains approximately the same. The electric field in the channel decreases with decreasing photoionization efficiency for both polarities.

An increase in the photoionization rate leads to the opposite influence. In this case, we have an acceleration of the positive streamer and a deceleration of the negative streamer (figure 14). Positive streamers cannot propagate without seed electrons produced due to photoionization. Therefore, the increase in the photoionization rate favors the development of the positive streamer.

Here, electron avalanches propagate to the ionization front of the streamer (figure 15(a)). For negative polarity, seed electrons originate avalanches that develop away from the head. This causes an increase in the head radius and a steeper electric field decrease with the distance from the head (figure 15(b)). Thus the effect of the production of seed electrons in front of the head depends on polarity solely due to different directions of avalanche propagation from the points of seed electron generation.

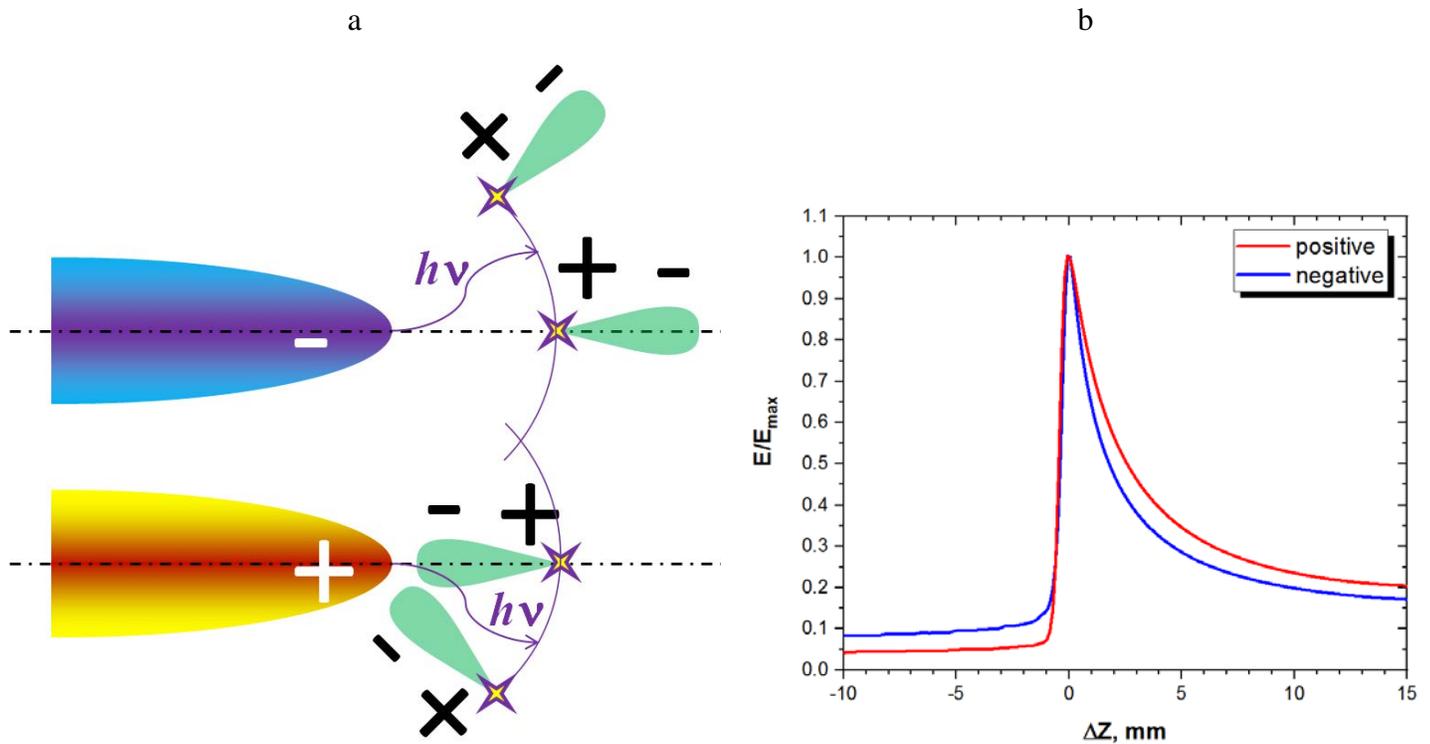

Figure 15. Scheme of processes in the vicinity of the heads of negative and positive streamers (a) and axial profiles of the normalized electric field in these regions (b). Normalization is to the maximum values of the electric field.

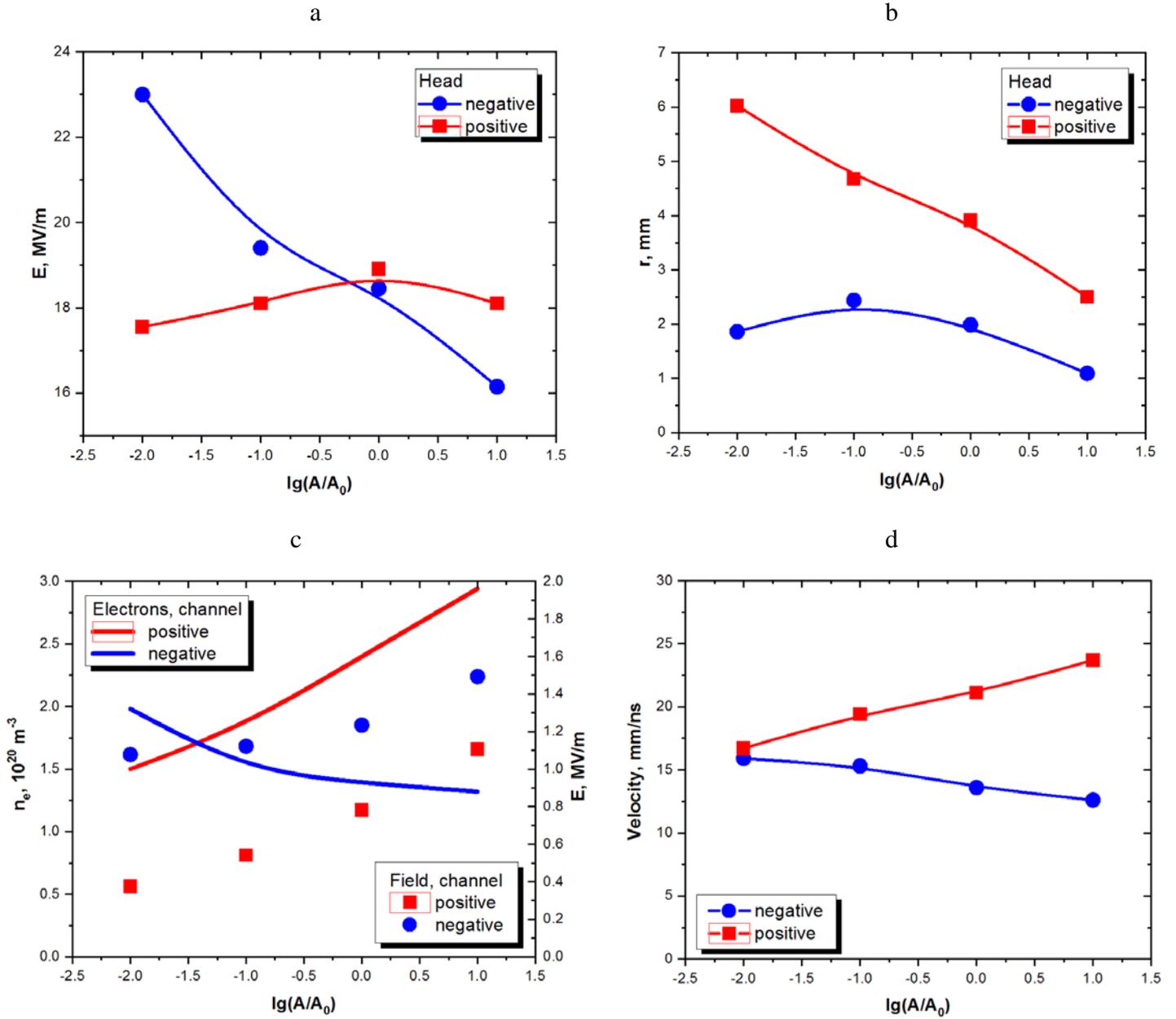

Figure 16. (a) Electric field at the head, (b) head radius, (c) electron density and electric field in the channel, and (d) streamer velocity versus the relative photoionization rate. Calculations are for $U = 300$ kV and streamer head position $Z_{head} \approx 80$ mm.

Figure 16 shows the calculated streamer characteristics as a function of the relative efficiency of photoionization, $A/A_0$, at the instant when the streamer head is at a distance of $Z_{head} \approx 80$ mm from the high-voltage electrode. Here, $A$ is the photoionization rate and $A_0$ is the ordinary photoionization rate in atmospheric pressure air. The curvature radius of the head of the negative streamer changes with the efficiency of photoionization only slightly. However, due to a change in the shape of the head, the electric field in the streamer head rises with decreasing photoionization rate for negative polarity. The shape of the streamer head for positive polarity is close to a hemispherical one. In this case, the electric field at the streamer head changes slightly, although the head radius increases with decreasing photoionization rate.

The variation in the photoionization efficiency affects the streamer velocity and electron density in the channel for the positive and negative streamers in different ways (figures 16(c) and 16(d)). For negative polarity, the velocity and electron density in the channel decrease with increasing rate of photoionization because of a decrease in the peak electric field at the streamer head. On the contrary, an increase in the photoionization rate causes an acceleration of the positive streamer and an increase in the electron density in the channel. The electric field in the streamer channel rises with increasing photoionization rate for both polarities.

This field depends on a number of the streamer parameters controlled by the electric field at the head. These parameters are (i) the plasma conductivity and electron density in the channel, (ii) the effective cross section of the channel carrying the discharge current, and (iii) the streamer velocity that governs the discharge current flowing through the channel to charge its new segments. As a result, the electric field in the channel of the positive streamer rises with increasing photoionization rate due to an increase in the streamer velocity and a decrease in the channel radius. For negative polarity, an increase in the electric field in the channel at higher photoionization rates is explained by the decreasing electron density in the channel (figure 16).

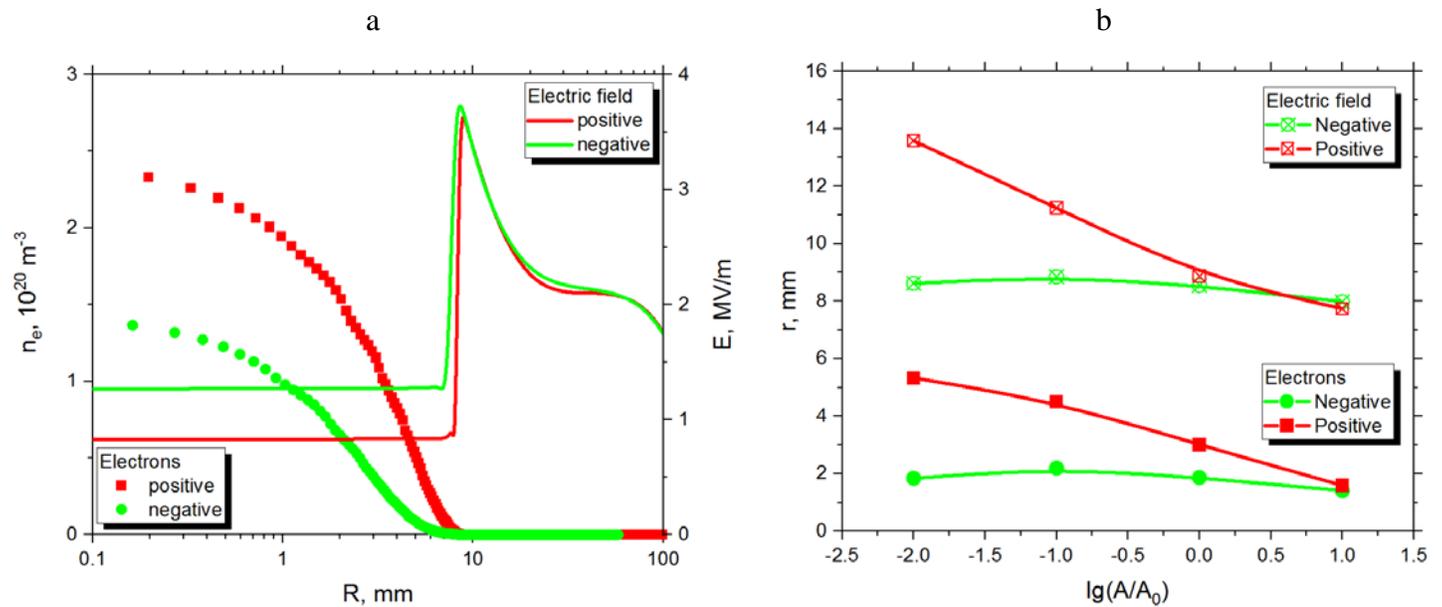

Figure 17. (a) Radial profiles of electric field and electron density and (b) channel radius determined from radial profiles of $E$ and $n_e$ versus the relative photoionization rate. Curves correspond to $t = 7$ ns for positive polarity and $t = 9.1$ ns for negative polarity (as in figure 14). Calculations are for $U = 300$ kV and $Z \approx 80$ mm.

Figure 17(a) demonstrates the radial profiles of the electric field and electron density in the cross section of the streamer channel. The position of the streamer head at this moment was $Z_{head} \approx 132$ mm for both the positive and negative streamers. The profiles were taken in the region of the maximum channel radius at $Z \approx 80$ mm. It is seen that the electron density profile is much narrower than the electric field profile. This is due to the distribution of the electric field at the streamer head (figure 3), which has a sharp maximum on the axis. In this case, the ionization and photoionization rates on the discharge axis are much higher than on the periphery, which leads to an inhomogeneous ionization over the streamer cross section. At the same time, the peak radial electric field is displaced from the axis due to the radial ionization wave, which develops as long as the electric field exceeds the breakdown threshold. In the end phase of the radial ionization wave propagation, ionization is not efficient. As a result, the channel radius corresponding to the peak radial electric field always exceeds the radius determined from the half-width of the electron density profile (figure 17(b)). Such a difference can be of fundamental importance in the analysis of the plasma instability development and discharge contraction. In our case, both the current radius (determined by the electron density) and electrodynamic radius (determined by the peak radial electric field) show qualitatively the same dependence on the photoionization rate (figure 17(b)). The channel radii of the negative streamer are practically independent of the photoionization rate, whereas the current and electrodynamic radii of the positive streamer decrease when the efficiency of photoionization increases.

It may be concluded that all differences between the positive and negative streamers follow from the opposite directions of electron avalanche development in front of the streamer head. The avalanches develop to the ionization front for positive voltage polarity and in the opposite direction for negative polarity (figure 15(a)). As a result, in the negative streamer, the avalanche development leads to an increase in the head radius, whereas the electric field at the head and in front of it (figure 15(b)), electron density in the channel, and streamer velocity, decrease. For positive polarity, the avalanches developing in front of the streamer head do not cause its expansion. Therefore, photoionization favors the enhancement of the local electric field in front of the head and faster propagation of the positive streamer in comparison with the negative streamer. A similar explanation of the difference between the properties of positive and negative streamers was given when discussing the simulated results in short air gaps [12]. In particular, it was mentioned that a larger field enhancement and higher velocity of positive streamers in short gaps are due to the lack of outward drift of seed electrons in front of the head. Evidently this is also true for long streamers.

## Simplified 1D analysis

To qualitatively analyze the results obtained, let us consider one-dimensional motion of electrons along the streamer axis in the electric field of a semi-spherical streamer head. In this model, the radius of the streamer head coincides with the radius of the streamer channel and the relation between the streamer velocity $v_s$, ionization frequency $v_i$ and the streamer radius $r_m$ is expressed as [1]

$$v_s = \frac{v_{im} r_m}{(2k-1) ln(n_m/n_0)}, \tag{8}$$

$$n_k \approx \frac{\varepsilon_0 v_{im}}{k e \mu_e}, \tag{9}$$

$$\frac{n_k}{n_m} \approx ln \frac{n_m}{n_0}, \tag{10}$$

where $v_{im}$ and $k$ are the parameters in the approximating expression for the ionization frequency $v_i(E) = v_{im}(E/E_m)^k$, $n_m$ is the electron density at the point of maximum electric field $E_m$ in the streamer head (see figures 10 and 11), $n_0$ is the density of seed electrons in front of the streamer head, $\mu_e$ is the electron mobility, $e$ is the elementary charge, and $\varepsilon_0$ is the permittivity of vacuum. The streamer radius $r_m$ is an input parameter in the model [1]. For positive streamers, this radius can be determined from the equation that depends on the ionization rate and photoionization rate [9]:

$$\frac{1}{2}\frac{p_q}{p+p_q}\int_{r_m}^{\infty}\Psi((z_0-r_m)p)exp\left(\int_{r_m}^{z_0}\frac{\alpha(z)v_d(z)}{v_s+v_d(z)}dz\right)dz_0 = 1 \ , \tag{11}$$

where $\alpha(z)$ is the first Townsend coefficient, $v_d(Z)$ is the electron drift velocity, $v_s$ is the streamer velocity, $p$ is the gas pressure, $p_q$ is the quenching pressure for the photo-ionizing states, and $\Psi((z_0-r_m)p)$ is the absorption coefficient of ionizing radiation.

To compare the streamer radius estimated from (11) and the calculated results, the head potential $U_{head}$ and streamer velocity are required. The potential $U_{head}$, as well as the streamer radius, determines the space distribution of electric field in front of the head:

$$E(z) = E_m \frac{r_m^2}{z^2} \tag{12}$$

for $z \geq r_m$ and $E(z) = 0$ for $z < r_m$. Here,

$$E_m \approx \frac{U_m}{2r_m} \tag{13}$$

In [9], equations (11)-(13) were used for positive streamers only because it was assumed that the properties of negative streamers are independent of photoionization in front of the streamer head and that the head is governed by other processes. However, as was shown in the previous section (equations (6) and (7)), photoionization plays a key role in the development of strong streamers for any polarity. In this case, equations (11)-(13) can be extended to the description of negative streamers if the total velocity $v_{st} + v_e(Z)$ in (11) is replaced by $v_{st} - v_e(Z)$. Then, equations (8)–(13) allow the formulation of a closed model to estimate all streamer parameters in the one-dimensional approximation for any polarity.

For the analytical 1D analysis, we take the streamer head potential and the streamer velocity from our 2D calculations. Then, equation (11) allows the calculation of head radius using the ionization rate, electron drift velocity and photoionization rate as input parameters. Contrary to the model [1] and equations (8)-(10), the model that is based on equation (11) does not contain the density of seed electrons (as an input parameter) since this density is directly calculated using the photoionization rate. Figure 18 compares the streamer radius calculated from equation (11) and the radii of the streamer head and channel determined in the 2D simulation for both voltage polarities. It follows from this comparison that the simple analytical model based on equation (11) qualitatively predicts the streamer radius in a wide range of applied voltages (from the thresholds of streamer initiation to the threshold of electrical breakdown in uniform gaps). In particular, the analytical 1D model shows a sharper rise of the streamer radius with the voltage increase for the positive streamer in comparison with the negative discharge, in qualitative agreement with the results of the 2D simulation. The difference between the streamer radii calculated from the 1D and 2D models is primarily associated with the assumption of constant radius of the streamer channel in the analytical approach [9]. In reality, the streamer channel extends during its development due to the radial ionizing wave (figure 9).

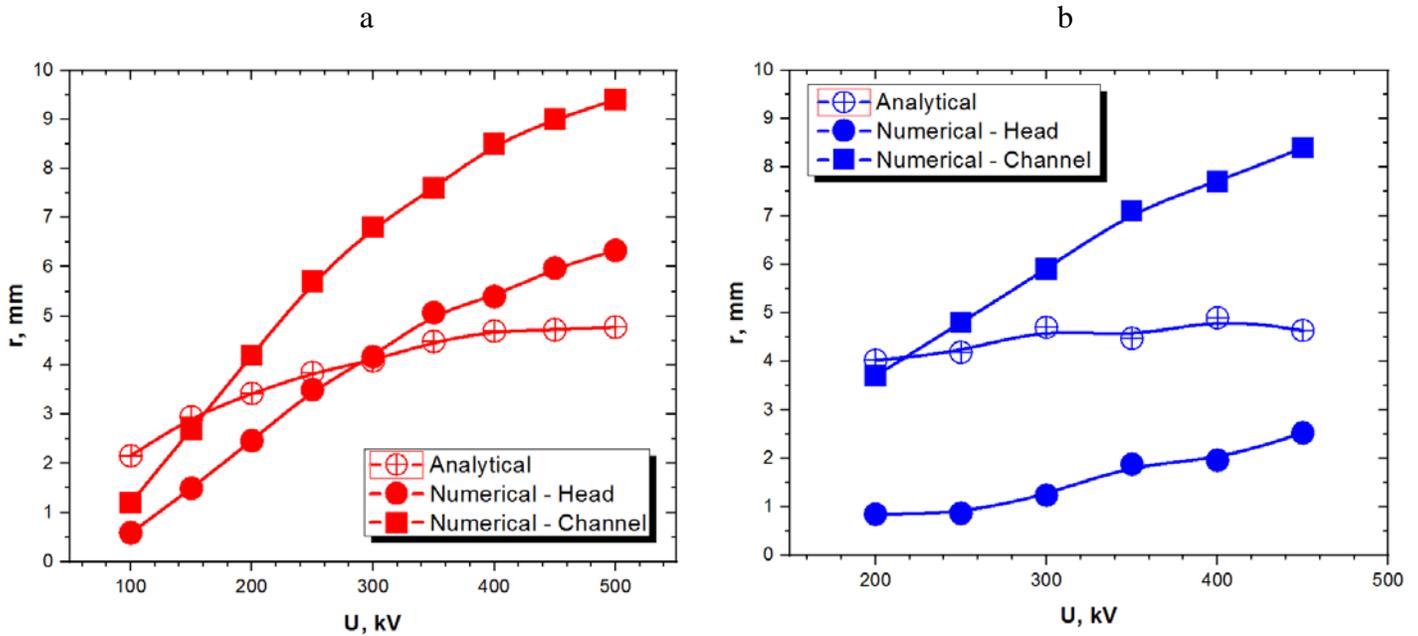

Figure 18. Radii of (a) positive and (b) negative streamers as a function of applied voltage. Calculations are made using the analytical 1D approach and numerical 2D approach for the instants when the streamer head is at $Z_{head} \approx 80$ mm.

To verify the conclusion about the importance of photoionization for streamer development, we analyze the dependence of the streamer radius on the photoionization efficiency using the simple 1D approach (equation (11)). Calculations are made for a voltage of 300 kV. The channel radius is calculated at the instant when the streamer head is at a distance of 80 mm from the high-voltage electrode. In the 1D calculation, the streamer velocity and the streamer head potential are taken from the 2D calculation as previously.

Figure 19 compares the channel radii obtained in both approaches for positive and negative polarities. There is good agreement between the 1D and 2D calculations. In particular, both models show that the channel radius is almost independent of the photoionization rate for the negative streamer. Here, according to the 2D calculation, the channel radius is around 5-6 mm when the rate of photoionization is varied by three orders of magnitude. In this case, the radius obtained from the 1D approach is around 5 mm. For the positive streamer, both models predict a 50% decrease in the channel radius when the photoionization rate increases by three orders of magnitude. In this case, it follows from the 1D model that the radius decreases from 6 to 4 mm, whereas the 2D model shows a decrease in the radius from 8 to 6 mm (figure 19). The difference between these data is within 30%.

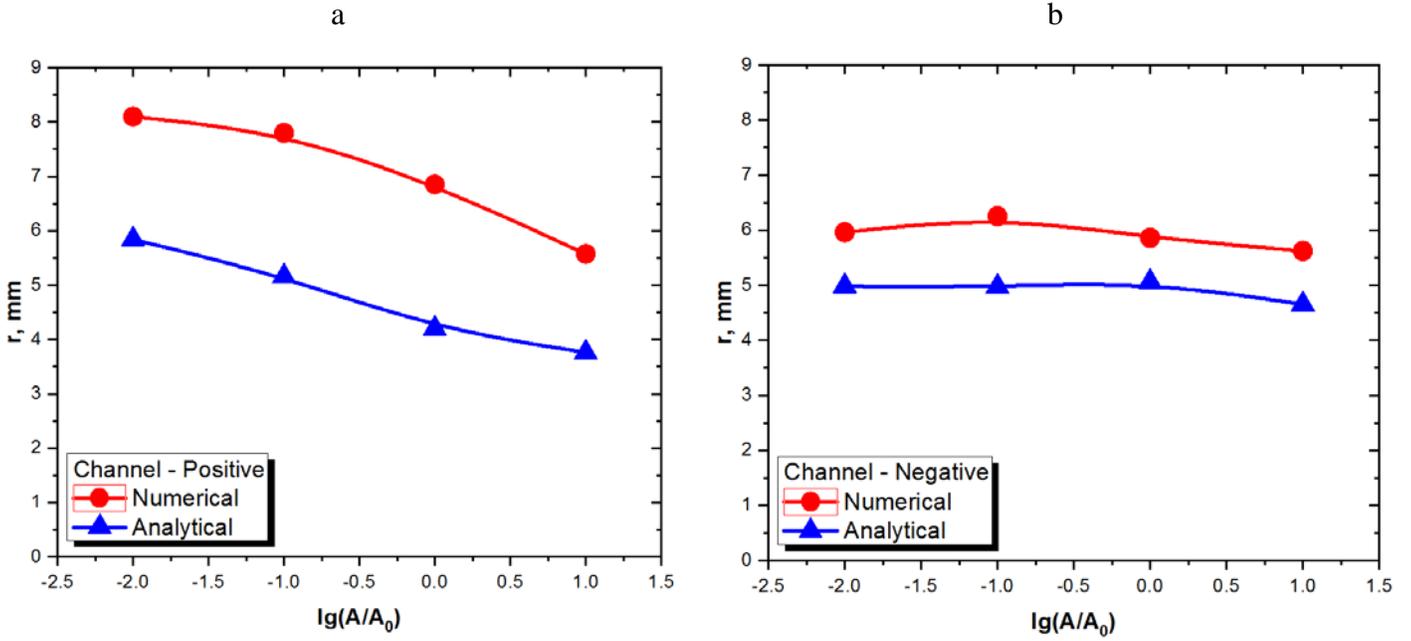

Figure 19. Radii of (a) positive and (b) negative streamer channels versus the relative photoionization rate. Calculations are made using the analytical 1D approach and numerical 2D approach for the instants when the streamer head is at $Z_{head} \approx 80$ mm.

Agreement between the predictions of the analytical 1D model and 2D simulation for the radius of the streamer channel allows us to use equations (8)-(11) for qualitative analysis of the effect of photoionization efficiency on other streamer characteristics (velocity and electron density in the channel). The channel radius of the negative streamer is almost independent of the photoionization rate. This means that the electric field at the head of the negative streamer depends only on the voltage drop across the channel. This voltage drop and the average electric field in the channel increase with increasing photoionization rate (figure 16(c)). As a result, the electric field at the head of the negative streamer should be lower for more efficient photoionization, in agreement with our 2D calculations (figure 16(a)). According to equations (8)-(10), a decrease in the electric field at the head leads to a sharp decrease in the streamer velocity because the electron-impact ionization frequency depends strongly ($v_i(E) \sim E^{2.5}$ [1]) on the electric field, whereas $r_m \approx$ const. Then, from (8)-(10), the velocity of the negative streamer should decrease with increasing photoionization rate, in agreement with the results of the 2D simulation shown in figure 16(d). It follows from (8)-(10) that a decrease in the electric field and electron-impact ionization frequency at the head causes a decrease in the electron density in the channel of the negative streamer. This also agrees with the 2D simulation of the negative streamer (figure 16(c)).

Let us consider similar characteristics for the positive streamer. Here, a decrease in the streamer radius with increasing photoionization efficiency (figure 19(a)) leads to an enhancement of the electric field at the head. As a result, according to (8)-(10), the streamer velocity and electron density in the channel are higher for higher photoionization rates. The 2D calculations confirm these predictions of the 1D model (figures 16(c) and

16(d)). It may be concluded that the simple 1D model based on equations (8)-(11) allows qualitative description of the properties of both positive and negative streamers in a wide range of the photoionization rate.

## Conclusions

The developed computer model predicts the properties of the positive and negative streamers in a long atmospheric pressure air gap. Calculations are made in a wide range of applied voltages, from the threshold of streamer initiation to the electrical breakdown in uniform air gaps. It is demonstrated that the peak electric field at the streamer head, velocity, electron density in the channel, discharge current, head radius and channel radius are higher for the positive streamer in comparison with the characteristics of the negative streamer. This result agrees with previous observations and calculations of the streamer velocity and radius in short air gaps. The difference between the properties of the positive and negative streamers follows from the opposite directions of electron avalanche development in front of the streamer head only. As a result, the electric field is more enhanced at the head of the positive streamer, by analogy with the processes in short streamers [12].

For all voltages studied, the average electric field in the channel of the negative streamer is approximately twice that in the channel of the positive streamer. This effect is explained by the difference in the characteristics (first of all, in the channel radius) of the negative and positive streamers. The obtained ratio between the values of the electric field in the channels of the negative and positive streamers agrees with numerous experimental observations in long air gaps. This difference allows bridging longer gaps by positive streamers and causes a stepped development of negative leaders with streamer zones in front of them. The stepped propagation of a negative leader discharge has been observed both in long laboratory gaps and during the development of downward lightning under thunderstorm conditions.

It is shown that streamer development critically depends on the generation of seed electrons by photoionization in front of the leading ionization wave for both voltage polarities. Variation in the photoionization rate leads to great and opposite in sign changes in the parameters of the positive and negative streamers.

The positive streamer fails to propagate without seed electrons in front of the head since the direction of the electron drift is opposite to the direction of the positive streamer propagation. Therefore, an increase of the amount of seed electrons with increasing photoionization rate leads to an increase in the streamer velocity and to a decrease in the curvature radius of the streamer head.

For the negative streamer, the radial development of electron avalanches causes a decrease in the electric field at the head and in the streamer velocity. The negative streamer becomes slower with increasing efficiency of photoionization. A decrease in the photoionization rate leads to a more efficient axial electron transport from the head of the negative streamer, as opposed to that of the positive streamer. As a result, the negative streamer tends to transform to an extremely thin cylindrical channel developing along the gap axis when photoionization rate is decreased.

A simple analytical 1D model is suggested and used to qualitatively describe the properties of the positive and negative streamers in long air gaps. The radius of the streamer channel and some other discharge characteristics are well predicted using this simplified approach in a wide range of photoionization rates.

## Acknowledgments

This work was supported by the Texas A&M University/DOE-Pittsburgh NETL project "Seed-Free MHD Topping Cycle for Coal and Gas Fired Power Generation".

## References

[1] Bazelyan E M and Raizer Yu P 1998 *Spark Discharge* (Boca Raton, Florida: CRC Press)

[2] Bazelyan E M and Raizer Yu P 2000 *Lightning Physics and Lightning Protection* (Bristol: IOP Publishing)

[3] Rakov V A and Uman M A 2003 *Lightning: Physics and Effects* (Cambridge: Cambridge University Press)

[4] Dwyer J R and Uman M A 2014 The physics of lightning *Phys. Rep.* **534** 147-241

[5] Fridman A 2008 *Plasma Chemistry* (Cambridge: Cambridge University Press)

[6] van Veldhuizen E M (ed) 2000 *Electrical Discharges for Environmental Purposes: Fundamentals and Applications* (New York: NOVA Science Publishers)

[7] Kong M G, Kroesen G, Morfill G, Nosenko T, Shimizu T, van Dijk J and Zimmermann J L 2009 Plasma medicine: an introductory review *New J. Phys.* **11** 115012

[8] Briels T M P, Kos J, Winands G J J, van Veldhuizen E M, and Ebert U 2008 Positive and negative streamers in ambient air: measuring diameter, velocity and dissipated energy *J. Phys. D: Appl. Phys.* **41** 234004

[9] Pancheshnyi S V, Starikovskaia S M, and Starikovskii A Yu 2001 Role of photoionization processes in propagation of cathode-directed streamer *J. Phys. D: Appl. Phys.* **34** 105–115

[10] Babaeva N Yu and Naidis G V 1997 Dynamics of positive and negative streamers in air in weak uniform electric fields *IEEE Trans. Plasma Sci.* **25** 375-379

[11] Naidis G V 2009 Positive and negative streamers in air: Velocity-diameter relation *Phys. Rev. E* **79** 057401

[12] Luque A, Ratushnaya V, and Ebert U 2008 Positive and negative streamers in ambient air: modelling evolution and velocities *J. Phys. D: Appl. Phys.* **41** 234005

[13] Liu N and Pasko V P 2004 Effects of photoionization on propagation and branching of positive and negative streamers in sprites *J. Geophys. Res.* **109** A04301

[14] Pasko V P 2007 Red sprite discharges in the atmosphere at high altitude: the molecular physics and the similarity with laboratory discharges *Plasma Sources Sci. Technol.* **16** S13–S29

[15] Luque A, Ebert U, Montijn C, and Hundsdorfer W 2007 Photoionization in negative streamers: Fast computations and two propagation modes *Appl. Phys. Lett.* **90** 081501

[16] Babaeva N Yu, Naidis G V, Tereshonok D V, and Son E E 2018 Development of nanosecond discharges in atmospheric pressure air: two competing mechanisms of precursor electrons production *J. Phys. D: Appl. Phys.* **51** 434002

[17] Nudnova M M, and Starikovskii A Yu 2008 Streamer head structure: role of ionization and photoionization. *J. Phys. D: Appl. Phys.* **41** 234003

[18] Pancheshnyi S V, Nudnova M M, and Starikovskii A Yu 2005 Development of a cathode-directed streamer discharge in air at different pressures: Experiment and comparison with direct numerical simulation. *Physical Review E.* **71** 016407

[19] Pancheshnyi S V and Starikovskii A Yu 2004 Stagnation dynamics of a cathode-directed streamer discharge in air. *Plasma Sources Sci. Technol.* **13** B1-B5

[20] Pancheshnyi S V and Starikovskii A Yu 2003 Two-dimensional numerical modeling of the cathode-directed streamer development in a long gap at high voltage *J. Phys. D: Appl. Phys.* **36** 2683-2691

[21] Zheleznyak M B, Mnatsakanyan A K, and Sizykh S V 1982 Photoionization of nitrogen and oxygen mixtures by radiation from a gas discharge *High Temp.* **20** 357-362

[22] Phelps C T and Griffiths R F 1976 Dependence of positive corona streamer propagation on air pressure and water vapor content *J. Appl. Phys.* **47** 2929-2934

[23] Kosarev I N, Starikovskiy A Yu, and Aleksandrov N L 2019 Development of high-voltage nanosecond discharge in strongly non-uniform gas *Plasma Sources Sci. Technol.* **28** 015005

[24] Starikovskiy A Yu and Aleksandrov N L 2019 'Gas-dynamic diode': Streamer interaction with sharp density gradients *Plasma Sources Sci. Technol.* **28** 095022

[25] Starikovskiy A Yu and Aleksandrov N L 2020 Blocking streamer development by plane gaseous layers of various densities. *Plasma Sources Sci. Technol.* **29** 034002